\documentclass[aps,prl,twoside,twocolumns,reprint,superscriptaddress]{revtex4-2}

\usepackage{mathtools}
\usepackage{graphicx}
\usepackage{amssymb}
\usepackage{amsmath}
\usepackage{amsthm}
\usepackage{amsfonts}
\usepackage{fancyvrb}
\usepackage{fancyhdr}
\usepackage{color}
\usepackage{xcolor}
\usepackage{braket}
\usepackage{soul}

\usepackage{enumerate}
\usepackage[utf8]{inputenc}
\usepackage{fancyhdr}
\usepackage{multirow}
\usepackage{color}
\usepackage{xcolor}
\usepackage{braket}
\usepackage{soul}
\usepackage{array}
\usepackage{subfigure}
\usepackage{afterpage}

\newcommand\blankpage{%
    \null
    \thispagestyle{empty}%
    \addtocounter{page}{-1}%
    \newpage}

\usepackage[printonlyused,withpage]{acronym}
\newacro{CCT}[CCT]{counterfactual concealed telecomputation}
\newacro{CUO}[CUO]{controlled unitary operation}
\newacro{DQC}[DQC]{distributed quantum computing}

\newtheoremstyle{mystyle}	
{}						
{}						
{\normalfont}				
{}						
{\bfseries}					
{:}						
{ }						
{ }						

\definecolor{CCTLABgreen}{RGB}{0,128,0}

\usepackage{bm}
\usepackage{graphicx}

\DeclareMathAlphabet{\mathsfbr}{OT1}{cmss}{m}{n}
\SetMathAlphabet{\mathsfbr}{bold}{OT1}{cmss}{bx}{n}
\DeclareRobustCommand{\msf}[1]{%
  \ifcat\noexpand#1\relax\msfgreek{#1}\else\mathsfbr{#1}\fi
}

\makeatletter
\newcommand{\msfgreek}[1]{\csname s\expandafter\@gobble\string#1\endcsname}
\makeatother

\DeclareFontEncoding{LGR}{}{} 
\DeclareSymbolFont{sfgreek}{LGR}{cmss}{m}{n}
\SetSymbolFont{sfgreek}{bold}{LGR}{cmss}{bx}{n}
\DeclareMathSymbol{\salpha}{\mathord}{sfgreek}{`a}
\DeclareMathSymbol{\sbeta}{\mathord}{sfgreek}{`b}
\DeclareMathSymbol{\sgamma}{\mathord}{sfgreek}{`g}
\DeclareMathSymbol{\sdelta}{\mathord}{sfgreek}{`d}
\DeclareMathSymbol{\sepsilon}{\mathord}{sfgreek}{`e}
\DeclareMathSymbol{\szeta}{\mathord}{sfgreek}{`z}
\DeclareMathSymbol{\seta}{\mathord}{sfgreek}{`h}
\DeclareMathSymbol{\stheta}{\mathord}{sfgreek}{`j}
\DeclareMathSymbol{\siota}{\mathord}{sfgreek}{`i}
\DeclareMathSymbol{\skappa}{\mathord}{sfgreek}{`k}
\DeclareMathSymbol{\slambda}{\mathord}{sfgreek}{`l}
\DeclareMathSymbol{\smu}{\mathord}{sfgreek}{`m}
\DeclareMathSymbol{\snu}{\mathord}{sfgreek}{`n}
\DeclareMathSymbol{\sxi}{\mathord}{sfgreek}{`x}
\DeclareMathSymbol{\somicron}{\mathord}{sfgreek}{`o}
\DeclareMathSymbol{\spi}{\mathord}{sfgreek}{`p}
\DeclareMathSymbol{\srho}{\mathord}{sfgreek}{`r}
\DeclareMathSymbol{\ssigma}{\mathord}{sfgreek}{`s}
\DeclareMathSymbol{\stau}{\mathord}{sfgreek}{`t}
\DeclareMathSymbol{\supsilon}{\mathord}{sfgreek}{`u}
\DeclareMathSymbol{\sphi}{\mathord}{sfgreek}{`f}
\DeclareMathSymbol{\schi}{\mathord}{sfgreek}{`q}
\DeclareMathSymbol{\spsi}{\mathord}{sfgreek}{`y}
\DeclareMathSymbol{\somega}{\mathord}{sfgreek}{`w}

\DeclareMathSymbol{\svarsigma}{\mathord}{sfgreek}{`c}

\DeclareMathSymbol{\sGamma}{\mathalpha}{sfgreek}{`G}
\DeclareMathSymbol{\sDelta}{\mathalpha}{sfgreek}{`D}
\DeclareMathSymbol{\sTheta}{\mathalpha}{sfgreek}{`J}
\DeclareMathSymbol{\sLambda}{\mathalpha}{sfgreek}{`L}
\DeclareMathSymbol{\sXi}{\mathalpha}{sfgreek}{`X}
\DeclareMathSymbol{\sPi}{\mathalpha}{sfgreek}{`P}
\DeclareMathSymbol{\sSigma}{\mathalpha}{sfgreek}{`S}
\DeclareMathSymbol{\sUpsilon}{\mathalpha}{sfgreek}{`U}
\DeclareMathSymbol{\sPhi}{\mathalpha}{sfgreek}{`F}
\DeclareMathSymbol{\sPsi}{\mathalpha}{sfgreek}{`Y}
\DeclareMathSymbol{\sOmega}{\mathalpha}{sfgreek}{`W}

\DeclareRobustCommand{\mcal}[1]{%
  \ifcat\noexpand#1\relax\mathnormal{#1}\else\cal{#1}\fi
}
\DeclareRobustCommand{\BM}[1]{%
  \ifcat\noexpand#1\relax\bm{\boldUppercaseItalicGreek{#1}}\else\bm{#1}\fi
}
\makeatletter
\newcommand{\boldUppercaseItalicGreek}[1]{\csname var\expandafter\@gobble\string#1\endcsname}
\makeatother
\newcommand{\rv}[1]{\msf{#1}}

\newcommand{\M}[1]{\BM{#1}}

\newcommand{\Prob}[1]{\mathbb{P}\hspace{-0.25ex}\left\{#1\right\}}

\DeclareMathVersion{timesmath}
\SetSymbolFont{letters}{timesmath}{OML}{ntxmi}{m}{it}
\DeclareMathVersion{timesmathbold}
\SetSymbolFont{letters}{timesmathbold}{OML}{ntxmi}{b}{it}

\newcommand{\slashslash}[1]{%
  \raisebox{#1}{%
    \scalebox{.7}{%
      \rotatebox[origin=c]{18}{$-$}%
    }%
  }%
}
\newcommand{\bslash}{%
  \mbox{%
   \vphantom{b}%
   \ooalign{\kern-.1em\smash{\slashslash{.9ex}}\hidewidth\cr$b$\cr}%
   \kern.05em
  }%
}
\newcommand{\slashed}[4]{%
  \mbox{%
   \vphantom{b}%
   \ooalign{\kern-#3\smash{\slashslash{#2}}\hidewidth\cr$#1$\cr}%
   \kern#4
  }}%
  
\newcommand{\bTRe}{\begin{dingautolist}{182}}
\newcommand{\eTRe}{\end{dingautolist}}

\begin{document}


\title{
	Counterfactual Concealed Telecomputation
}
\author{Fakhar Zaman}
\affiliation{Department of Electronics and Information Convergence Engineering, Kyung Hee University, Yongin-si, 17104 Korea}
\author{Hyundong Shin}
\thanks{Corresponding Author (hshin@khu.ac.kr).}
\affiliation{Department of Electronics and Information Convergence Engineering, Kyung Hee University, Yongin-si, 17104 Korea}
\author{Moe Z. Win}
\affiliation{Laboratory for Information and Decision Systems, Massachusetts Institute of Technology, Cambridge, MA 02139 USA}

\date{\today}

\begin{abstract}
Distributed computing is a fastest growing field---enabling virtual computing, parallel computing, and distributed storage. By exploiting the counterfactual  techniques, we devise a distributed blind quantum computation protocol to perform a universal two-qubit controlled unitary operation for any input state without using preshared entanglement and without exchanging physical particles between remote parties. This distributed protocol allows Bob to counterfactully apply an arbitrary unitary operator to Alice's qubit in probabilistic fashion, without revealing the operator to her, using a control qubit---called the \ac{CCT}. It is shown that the protocol is valid for general input states and that single-qubit unitary teleportation is a special case of \ac{CCT}. The quantum circuit for \ac{CCT} can be implemented using the  (chained) quantum Zeno gates and the protocol becomes determistic with simplified circuit implementation if the initial composite state of Alice and Bob is a Bell-type state.
\end{abstract}

\maketitle

A nonlocal \ac{CUO} is one of the fundamental building blocks in distributed quantum computing \cite{CEHM:99:PRA,DaiPenWin:J20b,CCTCGB:19:IEEEN} and quantum communications \cite{EJPP:00:PRA,FleShoWin:J08a,ChiConWin:J20}. Recently, it has been shown that any bipartite nonlocal unitary operation on a $d_\text{A} \times d_\text{B}$ dimensional quantum system can be implemented using at most $4d_\text{A}-5$ nonlocal controlled unitary operators, regardless of $d_\mathrm{B}$, where $d_\text{A}$ and $d_\text{B}$ denote the dimensions of quantum systems possessed by the remote parties Alice and Bob, respectively  \cite{CY:15:PRA}. In particular, a two-qubit  nonlocal controlled unitary operator plays an important role in distributed quantum computing as any $n$-qubit nonlocal unitary operation can be decomposed into a product of two-qubit nonlocal \acp{CUO} and single-qubit operations \cite{CY:15:PRA,BBCDMSSSW:95:PRA}.  Two-qubit nonlocal \acp{CUO} have been implemented using entanglement-assisted local operations and classical communication (LOCC) \cite{CY:14:PRA,STM:11:PRL}.

In general, a two-qubit \ac{CUO} can be represented as $\M{U}_\mathrm{c}=\M{I}\otimes\ket{0}\bra{0}+\M{U}\otimes\ket{1}\bra{1}$, where $\M{I}$ is the single-qubit identity operator and $\M{U}$ is an arbitrary single-qubit unitary operator. To devise a two-qubit nonlocal \ac{CUO}, $\M{U}_\mathrm{c}$ can be further decomposed to \cite{STM:11:PRL,LBCDS:15:PRL}
\begin{align}
\M{U}_\mathrm{c}
	=
	\left(
		\M{A}_1
		\otimes
		\M{B}_1
	\right)
	\bigg(
		\sum_{kl}
		e^{\iota kl\theta}
		\ket{kl}
		\bra{kl}
	\bigg)
	\left(
		\M{A}_2
		\otimes
		\M{B}_2
	\right),
\end{align}
where  $\iota=\sqrt{-1}$; $\ket{k}$ and $\ket{l}$ denote the computational basis of target and control qubits; and $\M{A}_1,\M{A}_2,\M{B}_1$ and $\M{B}_2$ are the single-qubit local unitary operators that depend on $\M{U}$. To date, nonlocal \acp{CUO} have been implemented using LOCC when the following two conditions are met: i) a sufficient amount of preshared entanglement is available and ii) $\M{U}$ is known to \emph{both} Alice and Bob \cite{STM:11:PRL}.

Blind quantum computation using entanglement-assisted LOCC \cite{GMMR:13:PRL,PF:15:PRL,F:17:NPJQI} is a unique capability enabled by quantum mechanics, which allows one party to use quantum computational resources of a remote party without revealing the input, computation, and output.  This quantum task enables a client to perform  computation remotely and privately under the unconditional security, which relies on the law of quantum mechanics such as quantum entanglement, quantum nonlocality and quantum nocloning theorem.
Counterfactual quantum communication \cite{SLAZ:13:PRL,AV:19:PRA} is an other unique capability enabled by quantum mechanics, which allows remote parties to communicate information without exchanging physical particles. The counterfactuality was first introduced based on the bomb detection experiment to determine the presence of a bomb in an interferometer without intreacting with it---namely, the \emph{interaction-free measurement}---by using the Mach-Zehnder  interferometer \cite{VE:93:FOP} and the quantum Zeno (QZ) effect \cite{KWHZK:95:PRL}. The direct counterfactual quantum communication protocol is based on the nested version of the Mach-Zehnder interferometer with $M$ outer and $N$ inner cycles---namely, the \emph{chained QZ (CQZ) effect}---where a classical bit is encoded in the the presence or absence of the absorptive object (e.g., shutter) in the interferometer. To ensure the counterfactuality of the protocol for both classical bits 0 and 1, if any physical particle is found in the quantum channel, the absorptive object absorbs the particle and declares the erasure of the classical information. In contrast, the quantum absorptive object (e.g., electron, trapped ions), which can take superposition of the absence and presence, is required to transfer quantum information in counterfactual way \cite{ZSW:19:arXiv}.
In addition to quantum communication, the counterfactuality has been successfully used in quantum computation \cite{HRBPK:06:Nature,KJHWKSJD:15:PRL} and quantum cryptography \cite{N:09:PRL,SH:14:PRA}.
\begin{figure*}[t!]
 \centering{
\includegraphics[width=0.925\textwidth]{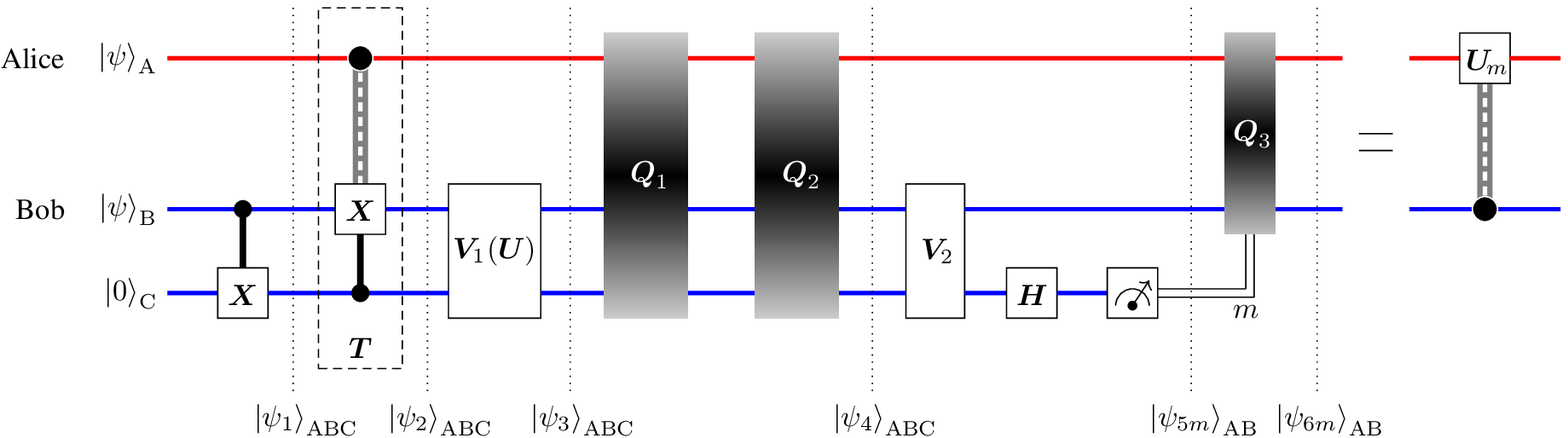}}
    \caption{
A \ac{CCT} protocol without preshared entanglement. Bob starts the protocol by entangling his qubit $\ket{\psi}_\text{B}$ and ancilla $\ket{0}_\text{C}$ with the local CNOT operation. Alice and Bob apply a sequence of nonlocal controlled flipping operations $\lbrace \M{T},\M{Q}_1,\M{Q}_2 \rbrace$ and the local operations $\lbrace\M{V}_1\hspace{-0.05cm}\left(\M{U}\right),\M{V}_{2}\rbrace$ (at Bob's side) where $\M{T}$ is a counterfactual Toffoli gate. At the end of the protocol, Bob applies the Hadamard gate $\M{H}$ on the ancilla followed by measuring the ancilla in the computational basis. Bob announces his measurement result $m\in\lbrace0,1\rbrace$ to Alice by using classical communication. Alice and Bob counterfactually apply the operation $\M{Q}_3$ on their composite state and transform the composite state as   $\ket{\psi_{6m}}_\text{AB}=\gamma\left(\M{I}\ket{\psi}_\text{A}\right)\otimes\ket{0}_{\text{B}}+\delta\left(\M{U}_{\hspace{-0.03cm}m}\ket{\psi}_\text{A}\right)\otimes\ket{1}_{\text{B}}$ where $\M{U}_{\hspace{-0.03cm}m}=\M{R}_\mathrm{z}\hspace{-0.05cm}\left(\phi\right)\M{R}_\mathrm{y}\hspace{-0.05cm}\left(\left(-1\right)^m\theta\right)\M{R}_\mathrm{z}\hspace{-0.05cm}\left(\varphi\right)$. Here $\M{I}$ denotes the single-qubit identity matrix and $\M{X}$ is the Pauli $x$ operator.
    }
    \label{fig:BNCU}
\end{figure*}

This letter put forth a new type of blind quantum computation  without using preshared entanglement  and without exchanging physical particles between remote parties.  
A fudamental building block of blind quantum computation is a protocol that enables Bob to counterfactually apply an arbitrary unitary operator $\M{U}$ to Alice's qubit in probabilistic fashion, without revealing $\M{U}$ to Alice, using a control qubit.
This is accomplished by decomposing the two-qubit \ac{CUO} corresponding to $\M{U}$ into global controlled  flipping operations and local operations at the remote parties. The key features of this protocol are that i) the global controlled flipping operations are implemented in counterfactual way, and  ii) both the global operations and Alice’s local operations are implemented in a way that $\M{U}$ is concealed from Alice. This protocol is called \acf{CCT}. The counterfactual implementation of global controlled flipping operations using the QZ \cite{IHBW:90:PRA} and CQZ gates \cite{ZJS:18:SR,ZJS:19:SR,ZSW:19:arXiv} is given in the supplementary material where Alice's qubit acts as a quantum absorptive object to ensure the counterfactuality of the protocol \footnote{See Supplementary Material for counterfactual implementation of global operations.}.  The protocol is shown to be valid for general input states, and single-qubit unitary teleportation \cite{HXZDG:04:PRL,HVCP:01:PRA} can be seen as a special case of \ac{CCT}. If the composite state of Alice and Bob is a Bell-type state, the protocol becomes deterministic and the quantum circuit for \ac{CCT} can be simplified significantly.

\emph{The protocol.}--- In general, a single-qubit unitary operator $\M{U}$ can be represented as $\M{U}=\M{R}_\mathrm{z}\hspace{-0.05cm}\left(\phi\right)\M{R}_\mathrm{y}\hspace{-0.05cm}\left(\theta\right)\M{R}_\mathrm{z}\hspace{-0.05cm}\left(\varphi\right)$ where $\phi,~\theta,$ and $\varphi$ are the Euler angles, and the rotation matrices are given by 
\begin{align}
\M{R}_\mathrm{y}\hspace{-0.05cm}\left(\theta\right)
	&
	=
	\renewcommand*{\arraystretch}{1.6}
	\begin{bmatrix}
		\cos\left(\theta/2\right) 
		& 
		-
		\sin\left(\theta/2\right)\\
		\sin\left(\theta/2\right) 
		& 
		\hspace{0.35cm}\cos\left(\theta/2\right)
	\end{bmatrix},\\
\M{R}_\mathrm{z}\hspace{-0.05cm}\left(\varphi\right)
	&
	=
	\renewcommand*{\arraystretch}{1.5}
	\begin{bmatrix}
		e^{-\iota\varphi/2} 
		& 
		0\\
		0 
		& 
		e^{\iota\varphi/2}
	\end{bmatrix}.
\end{align} 

\begin{figure*}[t!]
 \centering{
\includegraphics[width=0.7\textwidth]{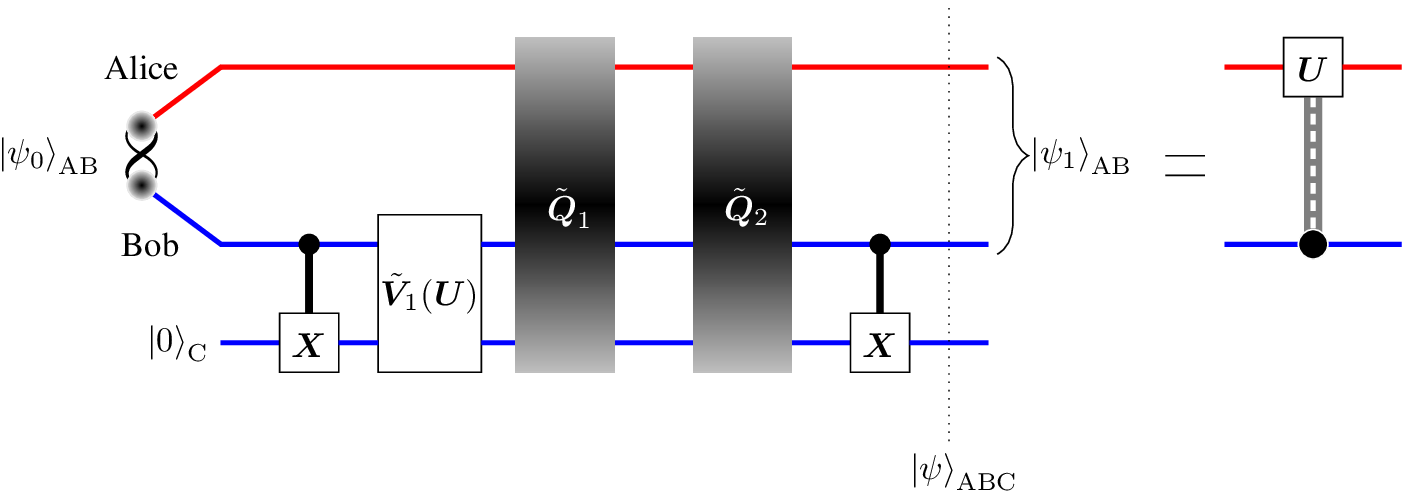}}
    \caption{
A deterministic \ac{CCT} protocol for Bell-type states to counterfactually apply an arbitrary unitary operator on the Alice's qubit in concealed and controlled fashion without using additional preshared entanglement. Similar to the \ac{CCT} protocol for general input states, Bob starts the protocol by entangling his qubit with ancilla $\ket{0}_\mathrm{C}$ with local CNOT operation. Alice and Bob apply a sequence of nonlocal controlled flipping operations $\lbrace \tilde{\M{Q}}_1,\tilde{\M{Q}}_2\rbrace$ and the local operation $\tilde{\M{V}}_1\hspace{-0.05cm}\left(\M{U}\right)$ (at Bob's side). At the end of the protocol, Bob applies the CNOT operation locally to disentangle the ancilla qubit and transforms the composite state as $\ket{\psi}_\text{ABC}=\big(\M{I}\otimes\ket{0}_\mathrm{B}\hspace{-0.065cm}\bra{0}+\M{U}\otimes\ket{1}_\mathrm{B}\hspace{-0.065cm}\bra{1}\big)\ket{\psi_0}_\mathrm{AB}\otimes\ket{0}_\mathrm{C}$ where $\M{U}=\M{R}_\mathrm{z}\hspace{-0.05cm}\left(\phi\right)\M{R}_\mathrm{y}\hspace{-0.05cm}\left(\theta\right)\M{R}_\mathrm{z}\hspace{-0.05cm}\left(\varphi\right)$. 
    }
    \label{fig:BTS-BNCU}
\end{figure*}
The proposed \ac{CCT} protocol will counterfactually implement controlled-$\M{U}_\rv{m}$ on Alice's qubit, without revealing $\M{U}_\rv{m}$ to Alice as shown in Fig.~\ref{fig:BNCU}, where 
\begin{align}
\M{U}_{\hspace{-0.03cm}\rv{m}}
	=
	\M{R}_\mathrm{z}
	\hspace{-0.05cm}
	\left(
		\phi
	\right)
	\M{R}_\mathrm{y}
	\hspace{-0.05cm}
	\left(
	\left(-1\right)^\rv{m}
		\theta
	\right)
	\M{R}_\mathrm{z}
	\hspace{-0.05cm}
	\left(
		\varphi
	\right),
\end{align}
with 
\begin{align}
\Prob{\rv{m}=m}
	=
	\begin{cases}
	\frac{1}{2}, &\text{for}~m=0,\\
	\frac{1}{2}, &\text{for}~m=1.
	\end{cases}
\end{align}
To demonstrate the implementation of \ac{CCT} (see Fig.~\ref{fig:BNCU}), consider the arbitrary pure input states of Alice's target qubit $\ket{\psi}_\text{A}$ and Bob's control qubit $\ket{\psi}_\text{B}$ as follows:
\begin{align}
\ket{\psi}_{\text{A}}
	&
	=
	\alpha
	\ket{0}_\text{A}
	+
	\beta
	\ket{1}_\text{A},\\
\ket{\psi}_\text{B}
	&
	=
	\gamma
	\ket{0}_\text{B}
	+
	\delta
	\ket{1}_\text{B},
\end{align}
with the complex coefficients $\alpha$, $\beta$, $\gamma$, and $\delta$ satisfying  $\vert\alpha\vert^2+\vert\beta\vert^2=1$ and $\vert\gamma\vert^2+\vert\delta\vert^2=1$. Bob starts the protocol by performing the local CNOT operation to entangle his qubit with the ancillary qutrit $\ket{0}_\mathrm{C}$. Alice and Bob conterfactually apply Toffoli gate $\M{T}$ by using the CQZ gate (see Fig.~S8 in the supplementary material),  with Alice's qubit and ancillary qutrit as control and Bob's qubit as a target. To ensure the conterfactuality of the protocol, if any physical particle is transmitted over the quantum channel, Alice's quantum absorptive object absorbs the particle and both parties (Alice and Bob) discard the protocol. Unless the protocol is discarded, Bob applies the following local operation $\M{V}_1\hspace{-0.05cm}\left(\M{U}\right)$ on his qubit and ancillary qutrit:
\begin{align}
\M{V}_1\hspace{-0.05cm}\left(\M{U}\right)
	=
	\M{V}_{14}
	\M{V}_{13}
	\M{V}_{12}
	\M{V}_{11},
\end{align} 
where 
\begin{align}
&
\begin{aligned}
\M{V}_{11}
	&=
	\M{I}
	\otimes
	\ket{0}_{\text{C}}
	\hspace{-0.065cm}
	\bra{0}
	+
	\left(
		\M{R}_\mathrm{z}\hspace{-0.05cm}\left(\varphi\right)		  				
		\M{X}
	\right)
	\otimes
	\ket{1}_{\text{C}}
	\hspace{-0.065cm}
	\bra{1}\\
	&\hspace{2.15cm}
	+
	\M{I}
	\otimes
	\ket{2}_{\text{C}}
	\hspace{-0.065cm}
	\bra{2}
\end{aligned}\\
&
\begin{aligned}
\M{V}_{12}
	&
	=
	\ket{0}_{\text{B}}
	\hspace{-0.065cm}
	\bra{0}
	\otimes
	\M{I}
	+
	\ket{10}_{\text{BC}}
	\hspace{-0.065cm}
	\bra{10}+
	\ket{12}_{\text{BC}}
	\hspace{-0.065cm}
	\bra{11}
	\\
	&\hspace{2.15cm}
	+
	\ket{11}_{\text{BC}}
	\hspace{-0.065cm}
	\bra{12},
\end{aligned}\\
&
\begin{aligned}
\M{V}_{13}
	&	
	=
	\M{I}
	\otimes
	\ket{0}_{\text{C}}
	\hspace{-0.065cm}
	\bra{0}
	+
	\left(
		\M{R}_\mathrm{z}\hspace{-0.05cm}\left(\phi\right)
		\M{R}_\mathrm{y}\hspace{-0.05cm}\left(\theta\right)
	\right)
	\otimes
		\ket{1}_{\text{C}}
		\hspace{-0.065cm}
		\bra{1}
		\\
		&\hspace{2.15cm}
		+
	\left(
		\M{R}_\mathrm{z}\hspace{-0.05cm}\left(\phi\right)
		\M{R}_\mathrm{y}\hspace{-0.05cm}\left(\theta\right)
	\right)
	\otimes
		\ket{2}_{\text{C}}
		\hspace{-0.065cm}
		\bra{2},
\end{aligned}
\\
&
\M{V}_{14}
	=
	\M{I}
	\otimes
	\left(
		\ket{0}_{\text{C}}
		\hspace{-0.065cm}
		\bra{0}
		+
		\ket{1}_{\text{C}}
		\hspace{-0.065cm}
		\bra{1}
	\right)
	+
	\M{X}
	\otimes
	\ket{2}_{\text{C}}
	\hspace{-0.065cm}
	\bra{2},
\end{align}
and $\M{X}$ denotes the Pauli $x$ operator. Note that the dependence of $\M{V}_1$ on $\M{U}$ is through $\M{V}_{11}$ and $\M{V}_{13}$. Now, Alice and Bob counterfactually apply the following two consecutive controlled flipping operations $\M{Q}_1$ and $\M{Q}_2$ by using the QZ and CQZ gates, respectively (see Fig.~S8 in the supplementary material) where
\begin{align}
&
\begin{aligned}
\M{Q}_1
	&
	=
	\M{I}
	\otimes
	\big(
		\ket{00}_{\text{BC}}
		\hspace{-0.065cm}
		\bra{00}
		+
		\ket{01}_{\text{BC}}
		\hspace{-0.065cm}
		\bra{01}
		\big.\\
		&\hspace{0.83cm}
		+
		\big.
		\ket{10}_{\text{BC}}
		\hspace{-0.065cm}
		\bra{10}
		+
		\ket{02}_{\text{BC}}
		\hspace{-0.065cm}
		\bra{02}
	\big)
	\\
	&\hspace{0.35cm}
	+
	\M{X}
	\otimes
	\big(
		\ket{11}_{\text{BC}}
		\hspace{-0.065cm}
		\bra{11}
		+
		\ket{12}_{\text{BC}}
		\hspace{-0.065cm}
		\bra{12}
	\big),
\end{aligned}
\\
&
\begin{aligned}
\M{Q}_2
	=
	\big(
		\ket{0}_{\text{A}}
		\hspace{-0.065cm}
		\bra{0}
		+
		\ket{1}_{\text{A}}
		\hspace{-0.065cm}
		\bra{1}
	\big)
	\otimes
	\M{I}
	&
	\otimes
	\ket{0}_{\text{C}}
	\hspace{-0.065cm}
	\bra{0}
	\\
	+
	\ket{1}_{\text{A}}
	\hspace{-0.065cm}
	\bra{1}
	\otimes
	\M{I}
	&
	\otimes
	\ket{1}_{\text{C}}
	\hspace{-0.065cm}
	\bra{1}
	\\
	+
	\ket{0}_{\text{A}}
	\hspace{-0.065cm}
	\bra{0}
	\otimes
	\M{I}
	&
	\otimes
	\ket{2}_{\text{C}}
	\hspace{-0.065cm}
	\bra{2}
	\\
	+
	\ket{0}_{\text{A}}
	\hspace{-0.065cm}
	\bra{0}
	\otimes
	\M{X}
	&
	\otimes
	\ket{1}_{\text{C}}
	\hspace{-0.065cm}
	\bra{1}
	\\
	+
	\ket{1}_{\text{A}}
	\hspace{-0.065cm}
	\bra{1}
	\otimes
	\M{X}
	&
	\otimes
	\ket{2}_{\text{C}}
	\hspace{-0.065cm}
	\bra{2}.
\end{aligned}
\end{align}
Bob applies the local operation $\M{V}_2$ on his qubit and ancillary qutrit:
\begin{align}
\begin{aligned}
\M{V}_{2}
	&
	=
	\ket{0}_{\text{B}}
	\hspace{-0.065cm}
	\bra{0}
	\otimes
	\M{I}
	\\
	&
	\hspace{0.35cm}+
	\ket{1}_{\text{B}}
	\hspace{-0.065cm}
	\bra{1}
	\otimes
	\big(
		\ket{0}_{\text{C}}
		\hspace{-0.065cm}
		\bra{1}
		+
		\ket{1}_{\text{C}}
		\hspace{-0.065cm}
		\bra{2}
		+
		\ket{2}_{\text{C}}
		\hspace{-0.065cm}
		\bra{0}
	\big),
\end{aligned}
\end{align}
followed by the Hadamard gate $\M{H}$ on the ancillary qutrit to disentangle the ancillary qutrit from Alice's and Bob's qubits. At the end of the protocol, Bob performs the measurement on the ancillary qutrit in the computational basis where $m\in\lbrace 0,1\rbrace$ is a measurement outcome with equal probability. Bob announces the measurement result with classical communication and applies the unitary operation $\M{Q}_3$ in counterfactual way where 
\begin{align}
\M{Q}_3
	=
	\begin{cases}
	\M{I}, &\text{for}~m=0,\\
	\left(
		\M{Z}
		\otimes
		\M{X}
	\right)
	\M{Z}_\mathrm{c}
	\left(
		\M{I}
		\otimes
		\M{X}
	\right), &\text{for}~m=1,
	\end{cases}
\end{align} 
and $\M{Z}_\mathrm{c}$ is the controlled-$\M{Z}$ operation and $\M{Z}$ denotes the Pauli $z$ operator. 
To implement the set of global flipping operations $\M{T}, \M{Q}_1,\M{Q}_2$ and $\M{Q}_3$ counterfactually, there is the nonzero probability---called the \emph{abortion rate} \footnote{See (S44) in Supplementary Material.}---that the physical particle is found in the quantum channel and the protocol fails in counterfactuality. This abortion rate vanishes asymptotically as the cycle numbers of QZ and CQZ gates increase. 
In case any physical particle is traveled over the quantum channel, Alice and Bob discard the protocol to ensure the full counterfactuality of \ac{CCT}. Unless the protocol is discarded, the \ac{CCT} protocol transforms the initial state $\ket{\psi_{0}}_{\text{AB}}=\ket{\psi}_\mathrm{A}\ket{\psi}_\mathrm{B}$ for the measurement outcome $m$ as follows: 
\begin{align}
\hspace{-0.192cm}
\ket{\psi_{6m}}_{\text{AB}}
	=
	\gamma
	\big(
		\M{I}
		\ket{\psi}_\text{A}
	\big)
	\otimes
	\ket{0}_{\text{B}}
	+
	\delta
	\big(
		\M{U}_{\hspace{-0.03cm}m}
		\ket{\psi}_\text{A}
	\big)
	\otimes
	\ket{1}_{\text{B}}.\label{eq:COT implementation}
\end{align}
The equation \eqref{eq:COT implementation} shows that Bob has successfully performed the \ac{CCT} on Alice's qubit. Note that the detailed state transformations $\ket{\psi_1}_\mathrm{ABC}$ to $\ket{\psi_{6m}}_\mathrm{AB}$ can be found in the supplementary material.

For the unitary teleportation as a special case of the \ac{CCT} protocol, Bob sets the initial state of his qubit  to $\ket{\psi}_{\mathrm{B}}=\ket{1}_\mathrm{B}$, then
\begin{align}
\ket{\psi_{6m}}_{\mathrm{AB}}
	=
	\big(
		\M{U}_{\hspace{-0.03cm}m}
		\ket{\psi}_{\mathrm{A}}
	\big)
	\otimes
	\ket{1}_\mathrm{B}. \label{eq: UGT}
\end{align}
Equation~\eqref{eq: UGT} shows that at the end of the protocol, the qubits of Alice and Bob are in a separable state, resulting in the unitary transformation of the arbitrary input state $\ket{\psi}_{\mathrm{A}}$ of Alice---also known as the \emph{quantum remote control}.

\emph{Bell-type states.}--- Consider that initial states of Alice and Bob are Bell-type states (see Fig.~\ref{fig:BTS-BNCU}). In general, the Bell-type states are given as 
\begin{align}
\ket{\psi_0}_\text{AB}:
	\begin{cases}
		\ket{\psi^\pm_{00}}_\text{AB}
			=
			\alpha
			\ket{00}_\text{AB}
			\pm
			\beta
			\ket{11}_\text{AB},\\
		\ket{\psi^\pm_{01}}_\text{AB}
			=
			\gamma
			\ket{01}_\text{AB}
			\pm
			\delta
			\ket{10}_\text{AB},
	\end{cases}
\end{align}
where $\ket{\psi^\pm_{0\ell}}_\text{AB}$,  $\ell=0,1$, are called the $\ell$-class states. Assume that Bob knows either the input is an $0$-class or $1$-class state. Similar to the general scheme, Bob starts the protocol by entangling his qubit with the ancillary qubit. Bob directly applies 
\begin{align}
\tilde{\M{V}}_1\hspace{-0.05cm}\left(\M{U}\right)
	=
	\M{I}
	\otimes
	\ket{0}_{\text{C}}
	\hspace{-0.065cm}
	\bra{0}
	+
	\big(
		\M{X}^{1-\ell}
		\M{U}
		\M{X}^{\ell}
	\big)
	\otimes
	\ket{1}_{\text{C}}
	\hspace{-0.065cm}
	\bra{1},
\end{align}
for the $\ell$-class states.
Now Alice and Bob counterfactually apply 
\begin{align}
\begin{aligned}
\tilde{\M{Q}}_1
	&
	=
	\M{I}
	\otimes
	\big(
		\ket{00}_{\text{BC}}
		\hspace{-0.065cm}
		\bra{00}
		+
		\ket{01}_{\text{BC}}
		\hspace{-0.065cm}
		\bra{01}
		\big.\\
		&\hspace{0.83cm}
		+
		\big.
		\ket{10}_{\text{BC}}
		\hspace{-0.065cm}
		\bra{10}
	\big)
	+
	\M{X}
	\otimes
	\big(
		\ket{11}_{\text{BC}}
		\hspace{-0.065cm}
		\bra{11}
	\big),
\end{aligned}
\end{align}
followed by 
\begin{align}
\begin{aligned}
\tilde{\M{Q}}_2
	=
	\big(
		\ket{0}_{\text{A}}
		\hspace{-0.065cm}
		\bra{0}
		+
		\ket{1}_{\text{A}}
		\hspace{-0.065cm}
		\bra{1}
	\big)
	\otimes
	\M{I}
	\otimes
	&
	\ket{0}_{\text{C}}
	\hspace{-0.065cm}
	\bra{0}
	\\
	+
	\ket{1}_{\text{A}}
	\hspace{-0.065cm}
	\bra{1}
	\otimes
	\M{X}^{1-\ell}
	\otimes
	&
	\ket{1}_{\text{C}}
	\hspace{-0.065cm}
	\bra{1}
	\\
	+
	\ket{0}_{\text{A}}
	\hspace{-0.065cm}
	\bra{0}
	\otimes
	\M{X}^{\ell}
	\otimes
	&
	\ket{1}_{\text{C}}
	\hspace{-0.065cm}
	\bra{1},
\end{aligned}
\end{align}
for the $\ell$-class states. 
At the end of the protocol, Bob applies the local CNOT operation at his qubits to disentangle the ancillary qubit. 
Again, there exists the abortion rate \footnote{See (S50) in Supplementary Material.} that the physical particle is transmitted over the quantum channel and the protocol fails to implement the set of global flipping operations $\tilde{\M{Q}}_1$ and $\tilde{\M{Q}}_2$ counterfactually---which tends to zero under the asymptotic limits.
In case any physical particle is traveled over the quantum channel, the protocol aborts to ensure the full counterfactuality. Unless the protocol is discarded,
the \ac{CCT} protocol for Bell-type states transforms $\ket{\psi_0}_\mathrm{AB}$ as follows:
\begin{align}
\ket{\psi}_\text{ABC}
	=
	\ket{\psi_1}_\mathrm{AB}
	\otimes
	\ket{0}_\text{C},
\end{align}
where
\begin{align}
\ket{\psi_1}_\mathrm{AB}
	=
	\big(
		\M{I}
		\otimes
		\ket{0}_\mathrm{B}
		\hspace{-0.065cm}
		\bra{0}
		+
		\M{U}
		\otimes
		\ket{1}_\mathrm{B}
		\hspace{-0.065cm}
		\bra{1}
	\big)
	\ket{\psi_0}_\mathrm{AB}.
\end{align}
As the ancillary qubit is already in a separable state with qubits $\ket{\psi_1}_\mathrm{AB}$, Bob does not need to perform measurement on his ancillary qubit.

In summary, we devise a new protocol for distributed quantum computing that allows Alice and Bob to apply a two-qubit \ac{CUO} on any input state in probabilistic fashion without using preshared entanglement, without revealing the unitary operator to Alice, and without exchanging physical particles between remote parties. 
\begin{itemize}
\item
As any $n$-qubit unitary operator can be decomposed into the product of two-qubit unitary operators and single qubit operators, the \ac{CCT} protocol enables Bob to apply an arbitrary unitary operation on Alice’s arbitrary number of qubits remotely in concealed way without transmitting any physical particle over the quantum channel and without using preshared entanglement.  
As nonlocal controlled flipping operators $\M{Q}_1$ and $\M{Q}_2$ (or $\tilde{\M{Q}}_1$ and $\tilde{\M{Q}}_2$) are independent of states, the protocol is oblivious to the input states of Alice and Bob, leading to universal concealed telecomputation on arbitrary input states. 

\item 
As the unitary teleportation is a special case of the \ac{CCT} protocol, Bob is able to appeal quantum remote control at a distinct party (Alice) without using transmitting any physical particle over the quantum channel and without using preshared entanelement.  
As unitary operator $\M{U}$ is unknown to Alice, the \ac{CCT} protocol can play an important role for the cryptography tasks such as quantum secure direct communication, quantum secret sharing and distributed quantum secure computation. While we do not discuss the security of the protocol, future work may consider suitable variations of the current protocol for the cryptography tasks.
\end{itemize}
The abortion rate of CCT decreases with the cycle numbers $M$ and $N$, whereas the stability of the nested Mach-Zehender interferometer improves with decreasing these cycle numbers. This problem may limit the success probability of the \ac{CCT} protocol and needs to be overcome in future work towards effective \ac{CCT}.

This work was supported in part by the the National Research Foundation of Korea under Grant  2019R1A2C2007037 and in part by the Office of Naval Research under Grant  N00014-19-1-2724.

\afterpage{\blankpage}
\pagebreak
\onecolumngrid
\begin{center}
\textbf{\large Supplementary Material: Counterfactual Concealed Telecomputation}
\end{center}
\setcounter{equation}{0}
\setcounter{figure}{0}
\setcounter{table}{0}
\setcounter{page}{1}
\makeatletter
\newtagform{supplementary}[S]()
\usetagform{supplementary}

\newcolumntype{P}[1]{>{\centering\arraybackslash}p{#1}}
\renewcommand{\thefigure}{S\arabic{figure}}
\renewcommand{\bibnumfmt}[1]{[S#1]}
\renewcommand{\citenumfont}[1]{S#1}
\renewcommand{\thefigure}{S\arabic{figure}}
\newenvironment{abstract}{
  \begin{center}%
    \bfseries\abstractname
  \end{center}}%
  {\vfill}

\newacro{QZ}[QZ]{quantum Zeno}
\newacro{CQZ}[CQZ]{chained QZ}
\newacro{IFM}[IFM]{interaction-free measurement}
\newacro{CEPI}[CEPI]{counterfactual electron-photon intereaction}
\newacro{DCEPI}[D-CEPI]{dual CEPI}
\newacro{DCFO}[DCFO]{distributed controlled flipping operation}
\newacro{DDCFO}[D-DCFO]{dual DCFO}

\begin{abstract}
We demonstrate the state transformations of the \ac{CCT} protocol followed by the counterfactual implementation of global operations without using preshared entanglement and without transmitting any physical particle over the quantum channel. We put forth i) the fundamental gates required for the counterfactual implementation of the global operations, ii) the \ac{CCT} protocol for general input states, and iii) the \ac{CCT} protocol for Bell-type input states.  We also obtain the probability that the protocol fails in  counterfactuality---called the abortion rate of CCT---which vanishes asymptotically.

\end{abstract}

\maketitle
\twocolumngrid

\section{Preliminaries}\label{sec: 1}
The counterfactual quantum communication \cite{SLAZ:13:PRL,AV:19:PRA} is based on the single-particle nonlocality and quantum measurement theory.  A quantum state usually collapses back to its initial state if the time between repeated measurements is short enough \cite{IHBW:90:PRA}. This \ac{QZ} effect  has been demonstrated to achieve the \ac{IFM}  where the the state of a photon acts as an unstable quantum state corresponding to the presence of the absorptive object \cite{KWHZK:95:PRL}. We begin with detailed state transformations of the \ac{CCT} protocol for general input states and then briefly review the overall actions of the \ac{QZ} and  \ac{CQZ} gates \cite{ZJS:18:SR,ZJS:19:SR} that are invoked to devise  \ac{CCT}.

\subsection{\ac{CCT}}

 Alice and Bob start the protocol with the initial composite state $\ket{\psi_0}_\mathrm{AB}=\ket{\psi}_\mathrm{A}\ket{\psi}_\mathrm{B}$ where $\ket{\psi}_\mathrm{A}$ and $\ket{\psi}_\mathrm{B}$ are given in (6) and (7) of the main text, respectively. Bob applies the CNOT operation locally where Bob's qubit acts as a control and the ancillary qutrit is a target (see Fig.~1 in the main text). It transforms $\ket{\psi_0}_\mathrm{AB}$ to 
\begin{align}
\ket{\psi_1}_\mathrm{ABC}
	=
	\ket{\psi}_\mathrm{A}
	\left(
		\gamma
		\ket{00}_\text{BC}
		+
		\delta
		\ket{11}_\text{BC}
	\right).\label{eq:psi 1}
\end{align} 

The counterfactual Toffoli gate $\M{T}$ transforms $\ket{\psi_1}_\mathrm{ABC}$ to
\begin{equation}
\begin{aligned}
\ket{\psi_2}_{\text{ABC}}
	&
	=
	\gamma
	\ket{\psi}_\mathrm{A}
	\ket{00}_\text{BC}
	\\
	&\hspace{0.35cm}+
	\delta
	\big(
		\alpha
		\ket{01}_\text{AB}
		+
		\beta
		\ket{10}_\text{AB}
	\big)
	\ket{1}_\text{C}.
\end{aligned}
\label{eq:psi 2}
\end{equation}
Then, the local operation $\M{V}_1\hspace{-0.05cm}\left(\M{U}\right)$ at Bob's side transforms
$\ket{\psi_2}_{\text{ABC}}$ to
%
\begin{align}
\begin{aligned}
\ket{\psi_3}_{\text{ABC}}
	&
	=
	\gamma
	\ket{\psi}_\mathrm{A}
	\ket{00}_\text{BC}
	\\
	&
	\hspace{0.35cm}
	+
	\delta
	\alpha 
	e^{-\iota\left(\varphi+\phi\right)/2}
	\cos\left(\theta/2\right)
	\ket{001}_{\text{ABC}}
	\\
	&\hspace{0.35cm}
	+
	\delta
	\alpha 
	e^{-\iota\left(\varphi-\phi\right)/2}
	\sin\left(\theta/2\right)
	\ket{011}_{\text{ABC}}
	\\
	&\hspace{0.35cm}
	+
	\delta
	\beta 
	e^{+\iota\left(\varphi+\phi\right)/2}
	\cos\left(\theta/2\right)
	\ket{102}_{\text{ABC}}
	\\
	&\hspace{0.35cm}
	-
	\delta
	\beta 
	e^{+\iota\left(\varphi-\phi\right)/2}
	\sin\left(\theta/2\right)
	\ket{112}_{\text{ABC}},
\end{aligned}
\label{eq:psi 3}
\end{align}
where $\phi$, $\theta$, and $\varphi$ are the Euler angles of the unitary operator $\M{U}=\M{R}_\mathrm{z}\hspace{-0.05cm}\left(\phi\right)\M{R}_\mathrm{y}\hspace{-0.05cm}\left(\theta\right)\M{R}_\mathrm{z}\hspace{-0.05cm}\left(\varphi\right)$.
\begin{figure}[t!]	
  \centering
\includegraphics[width=0.5\textwidth]{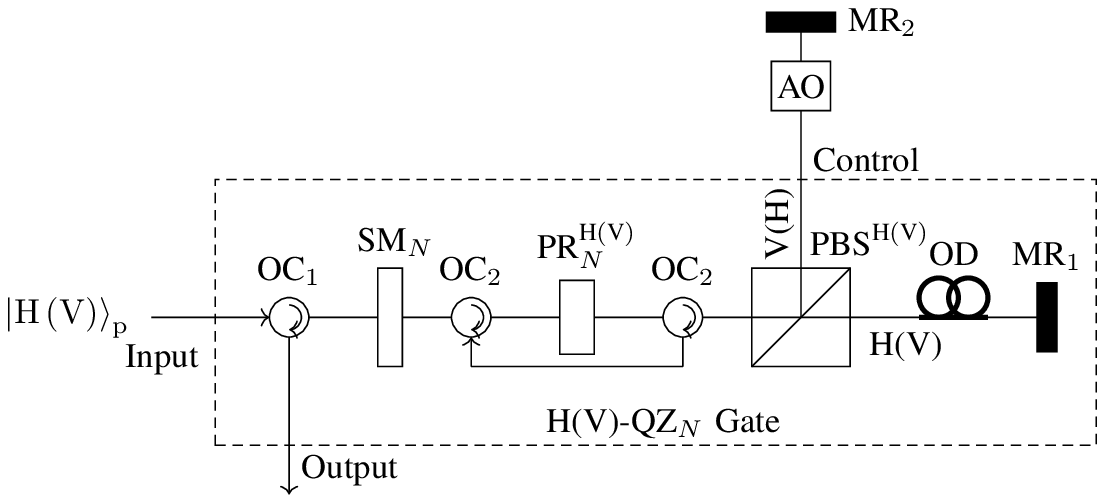}
    \caption{
     A H(V)-\ac{QZ}$_N$ gate with $N$ cycles where H (V) stands for horizontal (vertical) polarization of the photon, OC for an optical circulator, SM for a switchable mirror, PR for a polarizing rotator, PBS for a polarizing beam splitter, MR for a mirror, and AO shows the state of an absorptive object.
      }
    \label{fig:QZE}
\end{figure}
Now, Alice and Bob apply two counterfactual controlled flipping operations $\M{Q}_1$ and $\M{Q}_2$,
which transform the composite state $\ket{\psi_3}_{\text{ABC}}$ as follows:
\begin{equation}
\begin{aligned}
\ket{\psi_4}_{\text{ABC}}
	&
	=
	\gamma
	\ket{\psi}_\mathrm{A}
	\ket{00}_\text{BC}
	\\
	&\hspace{0.35cm}
	+
	\delta
	\alpha 
	e^{-\iota\left(\varphi+\phi\right)/2}
	\cos\left(\theta/2\right)
	\ket{011}_{\text{ABC}}
	\\
	&\hspace{0.35cm}
	+
	\delta
	\alpha 
	e^{-\iota\left(\varphi-\phi\right)/2}
	\sin\left(\theta/2\right)
	\ket{111}_{\text{ABC}}
	\\
	&\hspace{0.35cm}
	+
	\delta
	\beta 
	e^{+\iota\left(\varphi+\phi\right)/2}
	\cos\left(\theta/2\right)
	\ket{112}_{\text{ABC}}
	\\
	&\hspace{0.35cm}
	-
	\delta
	\beta 
	e^{+\iota\left(\varphi-\phi\right)/2}
	\sin\left(\theta/2\right)
	\ket{012}_{\text{ABC}}.
\end{aligned}
\label{eq:psi 4}
\end{equation}
\begin{table*}[t!]
\centering
\setlength{\tabcolsep}{8pt}
\renewcommand{\arraystretch}{1.5}
\caption{H(V)-\ac{QZ}$_N$ and H(V)-\ac{CQZ}$_{M,N}$ gates.}\label{tab:I-QZE}
\vspace{0.2cm}
\begin{tabular}{P{1.2cm} |  P{1.2cm} | P{1.2cm} | P{1.4cm} | P{2.4cm} | P{1.2cm} | P{1.4cm} | P{2.4cm}}
\hline
\multirow{2}{*}{Input} & \multirow{2}{*}{Control} &  \multicolumn{3}{c|}{\ac{QZ} Gate} & \multicolumn{3}{c}{\ac{CQZ} Gate}  \\
 \cline{3-8}
 & &  Output & Probability & Counterfactuality & Output & Probability & Counterfactuality\\
\hline
\hline 
\multirow{2}{*}{$\ket{\mathrm{H}\left(\mathrm{V}\right)}_{\mathrm{p}}$} & $\ket{0}_{\mathrm{AO}}$ &  $\ket{\mathrm{V}\left(\mathrm{H}\right)}_{\mathrm{p}}$ & 1 & No & $\ket{\mathrm{H}\left(\mathrm{V}\right)}_{\mathrm{p}}$ & $\lambda_0$ & Yes\\
& $\ket{1}_{\mathrm{AO}}$ &  $\ket{\mathrm{H}\left(\mathrm{V}\right)}_{\mathrm{p}}$ &  $\cos^{2N}\theta_N$ & Yes & $\ket{\mathrm{V}\left(\mathrm{H}\right)}_{\mathrm{p}}$ & $\lambda_1$ & Yes\\
\hline
\end{tabular}
\end{table*}
Bob's local operation $\M{V}_2$ followed by the Hadamard gate $\M{H}$ and the measurement on the ancillary qutrit in the computational basis collapses $\ket{\psi_4}_{\text{ABC}}$ to
\begin{align}
\label{eq:psi 5}
\ket{\psi_{5m}}_{\text{AB}}
	&
	=
	\gamma
	\ket{\psi}_\mathrm{A}
	\ket{0}_\text{B}\nonumber
	\\
	&\hspace{0.35cm}
	+
	\delta
	\alpha 
	e^{-\iota\left(\varphi+\phi\right)/2}
	\cos\left(\theta/2\right)
	\ket{01}_{\text{AB}}\nonumber
	\\
	&\hspace{0.35cm}
	+
	\delta
	\alpha 
	e^{-\iota\left(\varphi-\phi\right)/2}
	\sin\left(\theta/2\right)
	\ket{11}_{\text{AB}}
	\\
	&\hspace{0.35cm}
	+
	\left(-1\right)^{m}
	\delta	
	\beta 
	e^{+\iota\left(\varphi+\phi\right)/2}
	\cos\left(\theta/2\right)
	\ket{11}_{\text{AB}}\nonumber
	\\
	&\hspace{0.35cm}
	+
	\left(-1\right)^{1-m}
	\delta
	\beta 
	e^{+\iota\left(\varphi-\phi\right)/2}
	\sin\left(\theta/2\right)
	\ket{01}_{\text{AB}},\nonumber
\end{align}
where $m\in\lbrace 0,1\rbrace$ is the measurement outcome. Finally, the counterfactual global operation $\M{Q}_3$ transforms $\ket{\psi_{5m}}_{\text{AB}}$ for the measurement outcome $m$ as follows: 
\begin{align}
	\label{eq:psi 6}
\ket{\psi_{6m}}_{\text{AB}}
	&
	=
	\gamma
	\ket{\psi}_\mathrm{A}
	\ket{0}_\text{B}\nonumber
	\\
	&\hspace{0.35cm}
	+
	\delta
	\alpha 
	e^{-\iota\left(\varphi+\phi\right)/2}
	\cos\left(\theta/2\right)
	\ket{01}_{\text{AB}}\nonumber
	\\
	&\hspace{0.35cm}
	+
	\left(-1\right)^{m}
	\delta
	\alpha 
	e^{-\iota\left(\varphi-\phi\right)/2}
	\sin\left(\theta/2\right)
	\ket{11}_{\text{AB}}
	\\
	&\hspace{0.35cm}
	+
	\delta
	\beta 
	e^{+\iota\left(\varphi+\phi\right)/2}
	\cos\left(\theta/2\right)
	\ket{11}_{\text{AB}}\nonumber
	\\
	&\hspace{0.35cm}
	+
	\left(-1\right)^{1-m}
	\delta
	\beta 
	e^{+\iota\left(\varphi-\phi\right)/2}
	\sin\left(\theta/2\right)
	\ket{01}_{\text{AB}}, \nonumber
\end{align}
which can be written as (17) in the main text. 
Note that the probability $\zeta_m$ that the CCT protocol fails in counterfactuality---called the \emph{abortion rate}---is given in \eqref{eq:FP} 
when the global operations $\M{T},\M{Q}_1,\M{Q}_2$ and $\M{Q}_3$ are counterfactually implemented using the QZ and CQZ gates.

\subsection{\ac{QZ} Gates}

Fig.~\ref{fig:QZE} shows the Michelson version of the \ac{QZ} gate \cite{ZJS:19:SR} to perform \ac{IFM}. The \ac{QZ} gate is to ascertain the classical behavior of an absorptive object, i.e., to infer the absence state $\ket{0}_\mathrm{AO}$ or the presence state $\ket{1}_\mathrm{AO}$ of AO without interacting with it. The H(V)-\ac{QZ}$_N$ gate takes an H~(V) polarized photon as input. The switchable mirror SM$_N$ is initially turned off to allow passing the photon and is turned on for $N$ cycles once the photon is passed. After $N$ cycles, SM$_N$ is turned off again allowing the photon out.  The polarization rotator PR$_N^{\mathrm{H}\left(\mathrm{V}\right)}$ gives rotation to the input photon by an angle $\theta_N=\pi/\left(2N\right)$ as follows:
\begin{align}
\text{PR}_N^{\mathrm{H}\left(\mathrm{V}\right)}:
	\begin{cases}
	\ket{\mathrm{H}\left(\mathrm{V}\right)}_{\mathrm{p}}
		\rightarrow
		\cos\theta_N
		\ket{\mathrm{H}\left(\mathrm{V}\right)}_{\mathrm{p}}
		+
		\sin\theta_N
		\ket{\mathrm{V}\left(\mathrm{H}\right)}_{\mathrm{p}},\\
	\ket{\mathrm{V}\left(\mathrm{H}\right)}_{\mathrm{p}}
		\rightarrow
		\cos\theta_N
		\ket{\mathrm{V}\left(\mathrm{H}\right)}_{\mathrm{p}}
		-
		\sin\theta_N
		\ket{\mathrm{H}\left(\mathrm{V}\right)}_{\mathrm{p}}.
	\end{cases}
\end{align}
The photon state $\ket{\phi}$ after PR$_N^{\mathrm{H}\left(\mathrm{V}\right)}$ in the first cycle of the H(V)-\ac{QZ}$_N$ gate is given by
\begin{align}
\ket{\phi}=\cos\theta_N\ket{\mathrm{H}\left(\mathrm{V}\right)}_{\mathrm{p}}+\sin\theta_N\ket{\mathrm{V}\left(\mathrm{H}\right)}_{\mathrm{p}}. \label{eq:QZ pass}
\end{align}
Then, the polarizing beam splitter PBS separates the H and V components of the photon into two different optical paths: SM $\rightarrow$ MR$_1$ and SM $\rightarrow$ MR$_2$. The H~(V) component  goes towards MR$_1$ and the V~(H) component goes towards MR$_2$. The photon component in the second optical path only interacts with AO (control terminal). 
\begin{itemize}
\item 
$\text{AO}=\ket{0}_\mathrm{AO}$:
In the absence of the absorptive object, the V~(H) component of the photon is reflected by MR$_2$ and is returned back to PBS. Hence, the photon state remains unchanged. After $n \left(<N\right)$ cycles, the photon state is given by
\begin{align}	\label{eq:state:n}
\ket{\phi} 
	=
	\cos\left(n\theta_N\right)
	\ket{\mathrm{H}\left(\mathrm{V}\right)}_{\mathrm{p}}
	+
	\sin\left(n\theta_N\right)
	\ket{\mathrm{V}\left(\mathrm{H}\right)}_{\mathrm{p}}.
\end{align} 
The photon will end up in the state $\ket{\mathrm{V}\left(\mathrm{H}\right)}_{\mathrm{p}}$ with certainty by $\pi/2$ rotation after $N$ cycles.

\item 
$\text{AO}=\ket{1}_\mathrm{AO}$:
In the presence of the absorptive object, the V~(H) component is absorbed by AO if it is found in the control terminal. In each cycle, the probability of this absorption event is equal to $\sin^2\theta_N$. Unless the photon is absorbed, the photon state collapses to the initial state $\ket{\mathrm{H}\left(\mathrm{V}\right)}_{\mathrm{p}}$. 
After $N$ cycles, the photon is not absorbed and ends up in the state $\ket{\mathrm{H}\left(\mathrm{V}\right)}_{\mathrm{p}}$ with probability $\cos^{2N}\theta_N$ tending to one as $N\rightarrow\infty$.
\end{itemize}

Table~\ref{tab:I-QZE} shows the overall action of the \ac{QZ} gate. Note that the H(V)-\ac{QZ}$_N$ gate has the output $\ket{\mathrm{H}\left(\mathrm{V}\right)}_{\mathrm{p}}$ in the presence state $\ket{1}_\mathrm{AO}$  if the photon has not traveled over the control terminal (quantum channel). Hence, the \ac{QZ} gate is counterfactual only for this measurement outcome.

\begin{figure}[t!]
 \centering{
\includegraphics[width=0.35\textwidth]{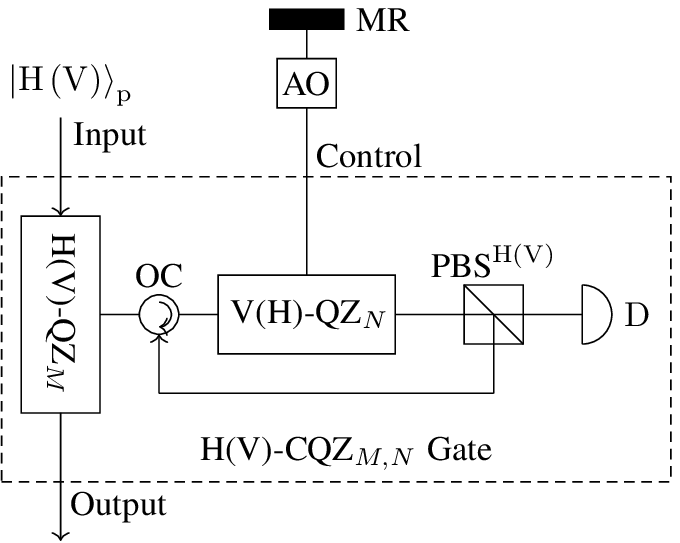}}
    \caption{
   A H(V)-\ac{CQZ}$_{M,N}$ gate with $M$ outer and $N$ inner cycles where D is a photon detector. Table~\ref{tab:I-QZE} shows the overall action of the H(V)-\ac{CQZ}$_{M,N}$ gate.    
   }
    \label{fig:MCQZ}
\end{figure}

\begin{figure*}[t!]
    \centering
    \subfigure[~Type I]
    {
        \includegraphics[width=0.4\textwidth]{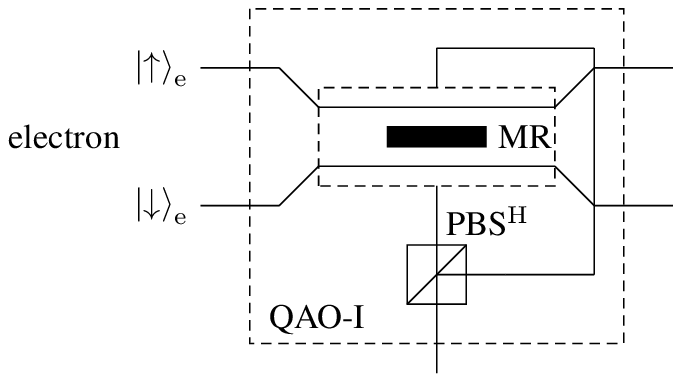}
        \label{fig:QAO:1}
    }
    \subfigure[~Type II]
    {
        \includegraphics[width=0.4\textwidth]{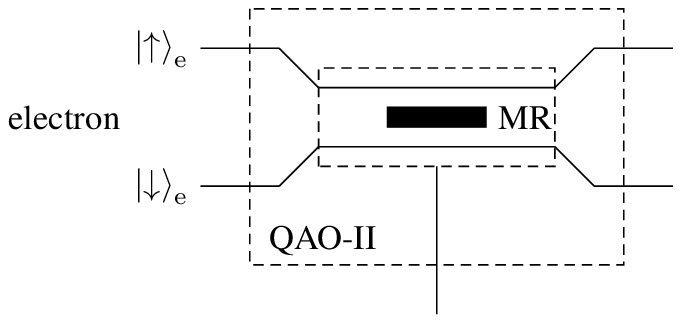}
        \label{fig:QAO:2}
    }
\caption{A quantum absorptive object (electron) for (a) the \ac{QZ} gate (type I) and (b) the \ac{CQZ} gate (type II). The electron takes the superposition of two paths $\ket{\uparrow}_{\mathrm{e}}$ and $\ket{\downarrow}_{\mathrm{e}}$. In type~I, the electron states
$\ket{\uparrow}_\mathrm{e}$ and $\ket{\downarrow}_\mathrm{e}$ act as the presence (absence) state $\ket{1 \left(0\right)}_\mathrm{AO}$ and the absence (presence) state $\ket{0 \left(1\right)}_\mathrm{AO}$ of the absorptive object for the H(V)-\ac{QZ} gate, respectively. In type~II, the electron states simply act as $\ket{\uparrow}_\mathrm{e}=\ket{0}_\mathrm{AO}$ and $\ket{\downarrow}_\mathrm{e}=\ket{1}_\mathrm{AO}$ for the \ac{CQZ} gate. If the photon is absorbed by the electron, the electron state is in an erasure state orthogonal to $\ket{\uparrow}_{\mathrm{e}}$ and $\ket{\downarrow}_{\mathrm{e}}$.
}
\label{fig:QAO}
\end{figure*}

\subsection{\ac{CQZ} Gates}
Fig.~\ref{fig:MCQZ} shows the nested version of \ac{QZ} gates with $M$ outer and $N$ inner cycles \cite{ZJS:18:SR}. The \ac{CQZ} gate enables to ascertain the absence or presence of the absorptive object counterfactually for both the outcomes. The H(V)-\ac{CQZ}$_{M,N}$ gate also takes an H~(V) polarized photon as input. In each outer cycle, the V~(H) component of the photon enters the inner V(H)-\ac{QZ}$_N$ gate.

\begin{itemize}

\item 
$\text{AO}=\ket{0}_\mathrm{AO}$:
In the absence of the absorptive object, the inner V(H)-\ac{QZ}$_N$ gate transforms the photon state $\ket{\mathrm{V}\left(\mathrm{H}\right)}_{\mathrm{p}}$ into $\ket{\mathrm{H}\left(\mathrm{V}\right)}_{\mathrm{p}}$ after $N$ cycles.
This component ends up at the detector D after PBS. Hence, the inner \ac{QZ} gate acts as an absorptive object for the outer \ac{QZ} gate in the absence state $\ket{0}_\mathrm{AO}$, where D serves to detect the event that the photon is found in the control terminal. In each outer cycle, unless the photon is discarded, the photon state collapses back to the initial state $\ket{\mathrm{H}\left(\mathrm{V}\right)}_{\mathrm{p}}$ with probability $\cos^2 \theta_M$. After $M$ outer cycles, the photon is not discarded at the detector D and ends up in the initial state $\ket{\mathrm{H}\left(\mathrm{V}\right)}_{\mathrm{p}}$ with probability 
\begin{align}
\lambda_0
	=
	\cos^{2M} \theta_M.
\end{align}
tending to one as $M \rightarrow \infty$.

\item 
$\text{AO}=\ket{1}_\mathrm{AO}$:
In case the absorptive object is present, the V~(H) component of the photon recombines with the H~(V) component and 
the photon state remains unchanged for the next outer cycle, unless the photon is absorbed by AO. 
Hence, the inner \ac{QZ} gate acts as a mirror for the outer \ac{QZ} gate in the presence state $\ket{1}_\mathrm{AO}$.  
After $i \left(<M\right)$ outer cycles, unless the photon is absorbed, the photon state is given by \eqref{eq:state:n}, 
which is again not absorbed by AO for the next outer cycle with probability 
\begin{align}
\left[
	1-
	\sin^2\left(i\theta_M\right)
	\sin^2\theta_N
\right]^{N}.
\end{align}
Hence, unless the photon is absorbed by AO, the H(V)-\ac{CQZ}$_{M,N}$ gate transforms the input state $\ket{\mathrm{H}\left(\mathrm{V}\right)}_{\mathrm{p}}$ into $\ket{\mathrm{V}\left(\mathrm{H}\right)}_{\mathrm{p}}$ with probability 
\begin{align}	\label{eq:P1}
\lambda_1
	=
	\prod_{i=1}^M
	\left[
		1-
		\sin^2\left(i\theta_M\right)
		\sin^2\theta_N
	\right]^{N}
\end{align}
tending to one as $M, N \rightarrow \infty$.

\end{itemize}

Note that the \ac{CQZ} gate is counterfactual for both the outcomes and infers the absence or presence of the absorptive object (with probability $\lambda_0$ or $\lambda_1$) but no physical particle (photon) is found in the control terminal (see Table~\ref{tab:I-QZE}).

\begin{figure}[t!]
 \centering{
\includegraphics[width=0.445\textwidth]{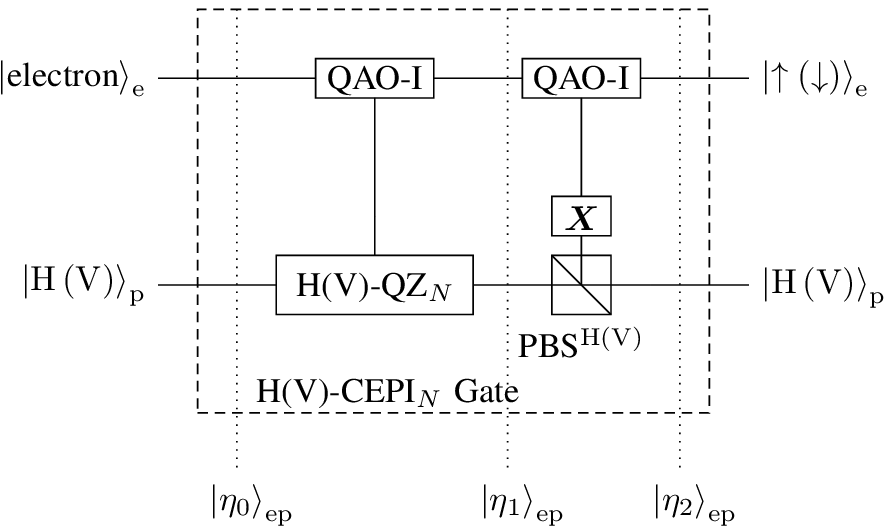}}
    \caption{
A H(V)-\ac{CEPI}$_N$ gate where the superposition state $\ket{\text{electron}}_\mathrm{e}=\alpha \ket{\uparrow}_{\mathrm{e}}+\beta\ket{\downarrow}_{\mathrm{e}}$ of the quantum absorptive object (electron) is collapsed to $\ket{\uparrow\left(\downarrow\right)}_\mathrm{e}$ using the H(V)-\ac{QZ}$_N$ gate unless the photon is absorbed by the electron. If the photon is found in the quantum channel, the pair of photon and electron is discarded in transforming $\ket{\eta_1}_\mathrm{ep}$ to $\ket{\eta_2}_\mathrm{ep}$ where the photon that has traveled over the channel is diverted again to the quantum absorptive object and absorbed by the electron. This electron-photon interaction is designed to output the photon and electron by using the presence state (blocking event) only. Hence, the protocol is fully counterfactual. 
   }
    \label{fig:H(V)-CEPI}
\end{figure}

\section{\ac{CCT} implementation using \ac{QZ} and \ac{CQZ} gates}\label{sec: 2}

As shown in Fig.~\ref{fig:QAO}, an electron as a quantum absorptive object takes superposition of two paths $\ket{\uparrow}_{\mathrm{e}}$ and $\ket{\downarrow}_{\mathrm{e}}$ where the subscript $\mathrm{e}$ denotes the electron. In type~I (Fig.~S\ref{fig:QAO:1}), the electron state $\ket{\uparrow\left(\downarrow\right)}_\mathrm{e}$ or $\ket{\downarrow\left(\uparrow\right)}_\mathrm{e}$ acts as the presence state $\ket{1}_\mathrm{AO}$ or the absence state $\ket{0}_\mathrm{AO}$ of the absorptive object for the H(V)-\ac{QZ}$_N$ gate.  For the counterfactuality of the protocol, we setup four gates:
\begin{enumerate}[1)]
\item
\acf{CEPI} gates using the H(V)-\ac{QZ}$_N$ gate;

\item
\ac{DCEPI} gates using  the dual \ac{QZ} (DQZ) gate;

\item 
\ac{DCFO} gates using the \ac{CEPI} gates;

\item
\ac{DDCFO} gates using the \ac{DCEPI} gates.
\end{enumerate} 
From the above list of gates, the second (first) and forth (third) gates are used to devise the \ac{CCT} for general (Bell-type) input states.

\subsection{\ac{CEPI} Gates}\label{sec: 2-1}

Fig.~\ref{fig:H(V)-CEPI} shows the H(V)-\ac{CEPI}$_N$ gate  where the quantum absorptive object is in the superposition state
\begin{align}
\ket{\text{electron}}_\mathrm{e}
=
	\alpha
	\ket{\uparrow}_{\mathrm{e}}
	+
	\beta
	\ket{\downarrow}_{\mathrm{e}}
	\label{eq:electron}
\end{align}
with $|\alpha|^2+|\beta|^2=1$.
Unless the photon is absorbed by the electron, the H(V)-\ac{CEPI}$_N$ gate collapses this quantum state by entangling and disentangling the electron-photon pair
\begin{align}
\ket{\eta_0}_\mathrm{ep}
=
	\ket{\text{electron}}_\mathrm{e}
	\ket{\mathrm{H}\left(\mathrm{V}\right)}_{\mathrm{p}}
\end{align} 
as follows:
\begin{align}	\label{eq:QHV:1}
\ket{\eta_0}_\mathrm{ep}
&\rightarrow
	\ket{\eta_1}_\mathrm{ep}
	=
	\alpha
	\ket{\uparrow\mathrm{H}}_{\mathrm{ep}}
	+
	\beta
	\ket{\downarrow\mathrm{V}}_{\mathrm{ep}} \\
&\rightarrow
	\ket{\eta_2}_\mathrm{ep}
	=
	\ket{\uparrow\left(\downarrow\right)}_\mathrm{e}
	\ket{\mathrm{H}\left(\mathrm{V}\right)}_{\mathrm{p}}
\end{align}
%
with probability
\begin{align}
\left(
	1-
	\nabla_0
	\sin^2\theta_N
\right)^N
\nabla_0,
\end{align}
where $\nabla_0=|\alpha|^2 \left(|\beta|^2\right)$ is the probability that the electron is in the presence state for the H(V)-\ac{QZ}$_N$ gate.

\begin{figure}[t!]
 \centering{
\includegraphics[width=0.475\textwidth]{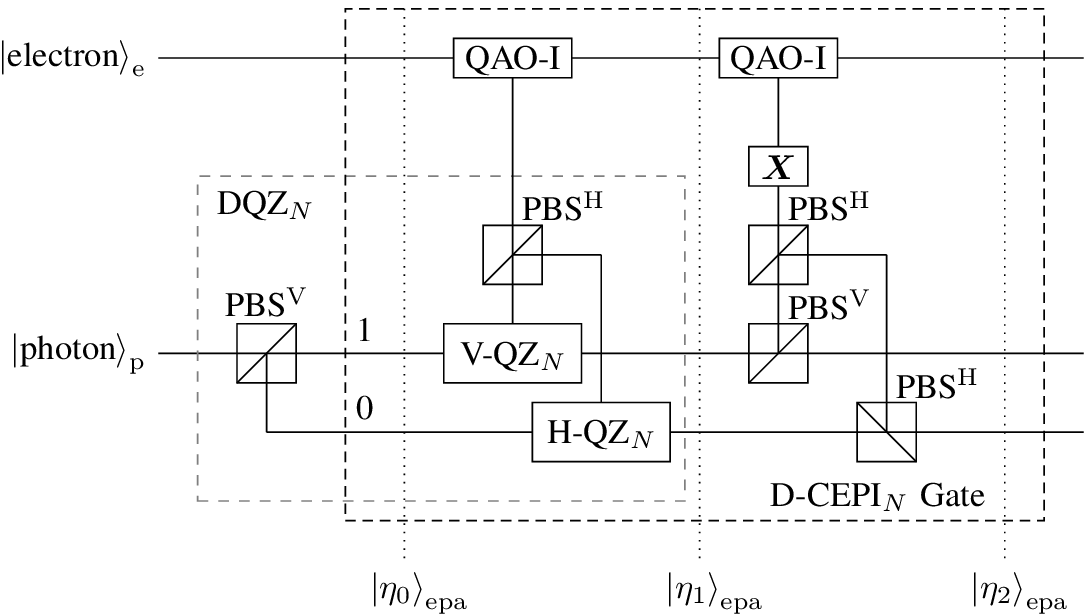}}
    \caption{
A \ac{DCEPI}$_N$ gate where the quantum absorptive object (electron) gets entangled with the existing photon (unless absorbed by the electron) using the DQZ$_N$ gate. Initially, the photon is in the superposition state $\ket{\text{photon}}_\mathrm{p}=\gamma\ket{0}_{\mathrm{p}}+\delta\ket{1}_{\mathrm{p}}$, which is entangled with the ancillary path state by PBS$^\mathrm{H}$ as $\ket{\text{photon}}_\mathrm{pa} = \gamma\ket{\mathrm{H0}}_{\mathrm{pa}} + \delta\ket{\mathrm{V1}}_{\mathrm{pa}}$ to start the \ac{DCEPI}. Similar to the H(V)-\ac{CEPI}$_N$ gate in Fig.~\ref{fig:H(V)-CEPI}, the \ac{DCEPI}$_N$ gate then transforms the electron-photon pair $\ket{\eta_0}_{\mathrm{epa}}$ to $\ket{\eta_2}_{\mathrm{epa}}=\gamma\ket{\uparrow\mathrm{H}0}_{\mathrm{epa}}+\delta\ket{\downarrow\mathrm{V}1}_{\mathrm{epa}}$ by using the blocking event only (unless the photon is absorbed by the electron).
}
    \label{fig:D-CEPI}
\end{figure}

The second (first) term of $\ket{\eta_1}_\mathrm{ep}$ is the outcome corresponding to the electron in the absence state for the H(V)-\ac{QZ}$_N$ gate. Since this outcome is not counterfactual, it is discarded (absorbed) by the electron using the PBS$^{\mathrm{H}\left(\mathrm{V}\right)}$ and the $\M{X}$ operator. To discard the factual (non-counterfactual) outcome $\ket{\mathrm{V}\left(\mathrm{H}\right)}_{\mathrm{p}}$ of the H(V)-\ac{QZ}$_N$ gate, PBS$^{\mathrm{H}\left(\mathrm{V}\right)}$ redirects this photon component to the quantum absorptive object (followed by the $\M{X}$ operator) to be absorbed by the electron. Hence, whenever the photon is found in the quantum channel, the electron absorbs it and becomes in an erasure state, leading the  H(V)-\ac{CEPI}$_N$ gate to output no photon and electron (e.g., particles in the erasure state).  This enables the protocol to abort nonlocally by discarding both the photon and the electron whenever its counterfactuality is broken.

\begin{figure*}[t!]
 \centering{
\includegraphics[width=0.85\textwidth]{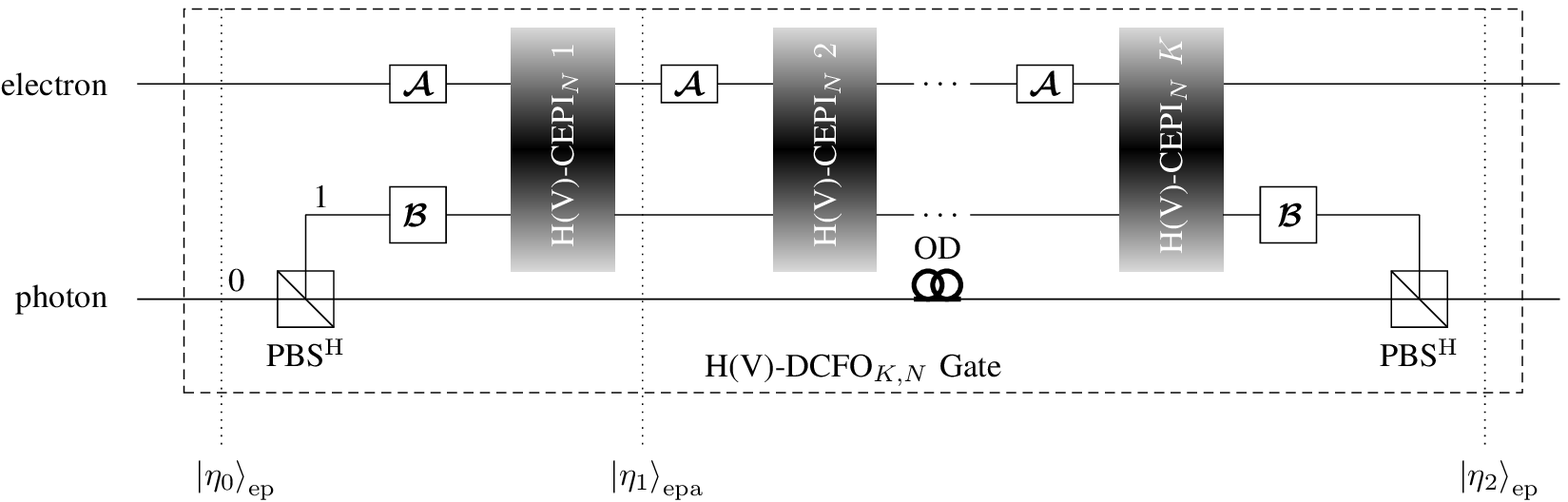}}
    \caption{
A H(V)-\ac{DCFO}$_{K,N}$ gate for Bell-type states of 1(0)-class.  Here $\M{\mathcal{A}}=\M{R}_\mathrm{y}\hspace{-0.05cm}\left(2\theta_K\right)$ is a rotation operation with rotation angle $\theta_K=\pi/\left(2K\right)$ and $\M{\mathcal{B}}=\M{X}^\ell$ for $\ell$-class states where $K$ is the number of \ac{CEPI} gates and $\M{R}_\mathrm{y}\hspace{-0.05cm}\left(2\theta_K\right)$ is the rotation around  $y$-axis.
}
    \label{fig:H(V)-DCFO}
\end{figure*}

\subsection{\ac{DCEPI} Gates}
Fig.~\ref{fig:D-CEPI} shows the dual form of the H(V)-\ac{CEPI}$_N$ gate in Fig.~\ref{fig:H(V)-CEPI}. For counterfacuality, this \ac{DCEPI}$_N$ gate works similarly to the H(V)-\ac{CEPI}$_N$ gate. The only difference is that the superposition polarization state
\begin{align}
\ket{\text{photon}}_\mathrm{p}=\gamma\ket{\mathrm{H}}_{\mathrm{p}}+\delta\ket{\mathrm{V}}_{\mathrm{p}}\label{eq:photon}
\end{align} 
of the input photon is entangled with the ancillary path state in the DQZ$_N$ gate as follows:  
\begin{align}
\ket{\text{photon}}_\mathrm{pa}
=
	\gamma\ket{\mathrm{H0}}_{\mathrm{pa}}
	+
	\delta\ket{\mathrm{V1}}_{\mathrm{pa}},	
\end{align}
where the ancilla states $\ket{0}_\mathrm{a}$ and $\ket{1}_\mathrm{a}$ show the paths for the H- and V-\ac{QZ} gates, respectively. Unless the photon is absorbed by the electron, the \ac{DCEPI}$_N$ gate transforms the electron-photon pair
\begin{align}
\ket{\eta_0}_\mathrm{epa}
=
	\ket{\text{electron}}_\mathrm{e}
	\ket{\text{photon}}_\mathrm{pa}
\end{align} 
as follows:
\begin{align}	
&\begin{aligned}
\ket{\eta_0}_\mathrm{epa}
\rightarrow
	\ket{\eta_1}_\mathrm{epa}
	&=
	\alpha
	\gamma
	\ket{\uparrow\mathrm{H}0}_{\mathrm{epa}}
	+
	\beta
	\gamma
	\ket{\downarrow\mathrm{V}0}_{\mathrm{epa}} 
	\\
	&\hspace{0.35cm}+
	\alpha
	\delta
	\ket{\uparrow\mathrm{H}1}_{\mathrm{epa}}
	+
	\beta
	\delta
	\ket{\downarrow\mathrm{V}1}_{\mathrm{epa}}
\end{aligned}\\
&~~~~~~~~~\rightarrow
	\ket{\eta_2}_\mathrm{epa}
	=
	\gamma
	\ket{\uparrow\mathrm{H}0}_{\mathrm{epa}}	
	+
	\delta
	\ket{\downarrow\mathrm{V}1}_{\mathrm{epa}}	
\end{align}
%
with probability 
\begin{align}
\left(
	1-
	\nabla_1
	\sin^2\theta_N
\right)^N
\nabla_1,
\end{align}
where
%
$\nabla_1
	=
	\vert
	\alpha
	\gamma
	\vert^2
	+
	\vert
	\beta
	\delta
	\vert^2
$ 
is the probability that the electron is in the presence state for \ac{QZ} gates in both paths.

\subsection{\ac{DCFO} Gates}
To devise the \ac{DCFO} gate, $K$ H(V)-\ac{CEPI}$_N$ gates are concatenated serially where Alice has the quantum absorptive object (electron) and Bob equips the \ac{QZ} gates (see Fig.~\ref{fig:H(V)-DCFO}). To explain the operation of the \ac{DCFO}$_{K,N}$ gate, consider  the Bell-type state of electron-photon pair  
\begin{align}
\ket{\eta_0}_\text{ep}:
	\begin{cases}
		\ket{\eta^\pm_{00}}_\text{ep}
			=
			\alpha\ket{\uparrow\mathrm{H}}_\text{ep}
			\pm
			\beta\ket{\downarrow\mathrm{V}}_\text{ep},\\
		\ket{\eta^\pm_{01}}_\text{ep}
			=
			\gamma\ket{\uparrow\mathrm{V}}_\text{ep}
			\pm
			\delta\ket{\downarrow\mathrm{H}}_\text{ep},\\
	\end{cases}\label{eq:H(V)-DCFO input}
\end{align}
where $\ket{\eta^\pm_{00}}_\text{ep}$ and $\ket{\eta^\pm_{01}}_\text{ep}$ are $0$- and $1$-class states, respectively. The H(V)-\ac{DCFO}$_{K,N}$ protocol for Bell-type states of 1(0)-class 
takes the following steps.

\begin{figure*}[t!]
 \centering{
\includegraphics[width=0.95\textwidth]{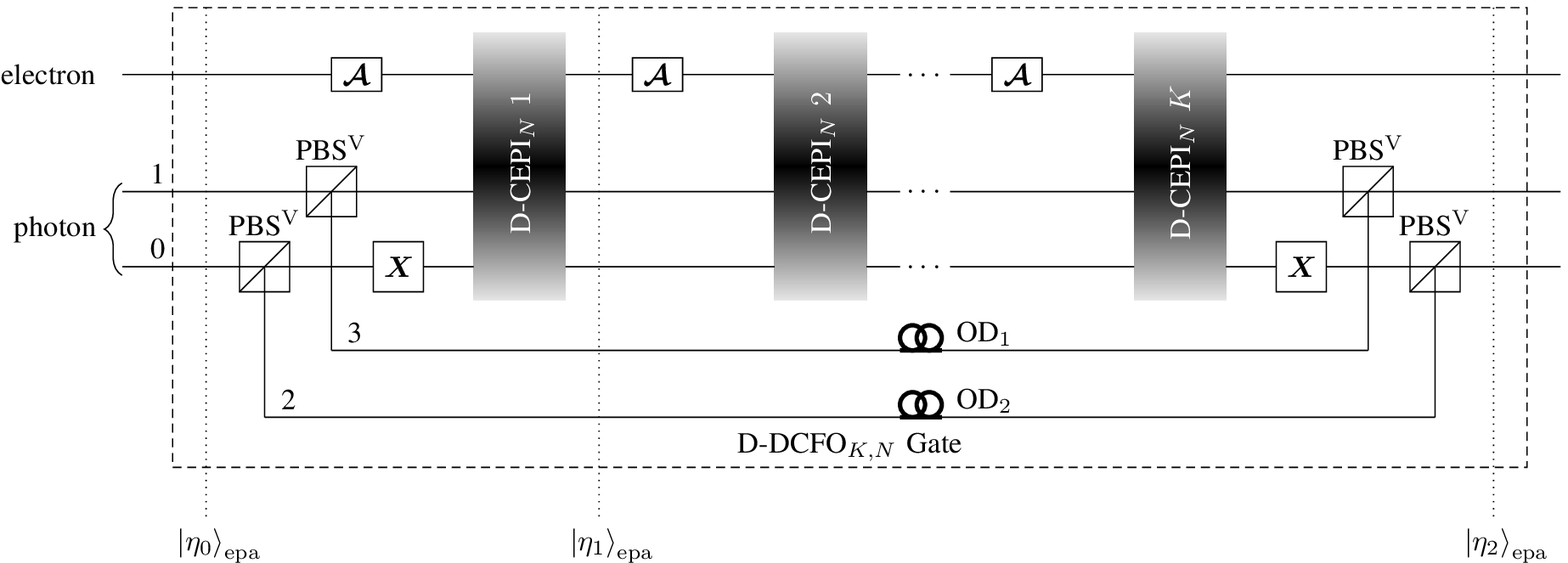}}
    \caption{
A dual version of the H(V)-\ac{DCFO}$_{K,N}$ gate. 
}
    \label{fig:D-DCFO}
\end{figure*}

\begin{enumerate}
\item
Bob starts the H(V)-\ac{DCFO}$_{K,N}$ protocol by throwing his photon towards PBS$^{\mathrm{V}}$, which allows the V component to pass and detour the H component to be recombined after $K$ successive H(V)-\ac{CEPI}$_N$ operations. Bob performs the $\M{\mathcal{B}}$ operator on the V photon component  where $\M{\mathcal{B}}=\M{X}^\ell$ for $\ell$-class states.

\item
Alice performs $\M{R}_\mathrm{y}\hspace{-0.05cm}\left(2\theta_K\right)$
on her qubit (electron) where $\theta_K=\pi/\left(2K\right)$. The rotation gate $\M{R}_\mathrm{y}\hspace{-0.05cm}\left(2\theta_K\right)$ transforms $\ket{\uparrow}_{\mathrm{e}}$ and $\ket{\downarrow}_{\mathrm{e}}$ as follows:
\begin{align}
\ket{\uparrow}_{\mathrm{e}}
	&
	\rightarrow
	\cos\theta_K
	\ket{\uparrow}_{\mathrm{e}}
	+
	\sin\theta_K
	\ket{\downarrow}_{\mathrm{e}},
	\\
\ket{\downarrow}_{\mathrm{e}}
	&
	\rightarrow
	\cos\theta_K
	\ket{\downarrow}_{\mathrm{e}}
	-
	\sin\theta_K
	\ket{\uparrow}_{\mathrm{e}}.
\end{align}

\item 
Bob inputs the H~(V) component to the H(V)-\ac{CEPI}$_N$ gate. 
Unless the photon is absorbed by the electron, the first H(V)-\ac{CEPI}$_N$ gate transforms the $\ket{\eta_0}_\text{ep}$ as follows: 
\begin{widetext}
\begin{align}
\ket{\eta_1}_\text{epa}:
	\begin{cases}
	\ket{\eta^\pm_{10}}_\text{epa}
	=
	\alpha
	\bigl(	
		\cos\theta_K
		\ket{\uparrow\mathrm{H}0}_\mathrm{epa}
		+
		\sin\theta_K
		\ket{\downarrow\mathrm{H}0}_\mathrm{epa}
	\bigr)
	\pm
	\beta
	\ket{\downarrow\mathrm{V}1}_\mathrm{epa},\\
	\ket{\eta^\pm_{11}}_\text{epa}
	=
	\gamma
	\ket{\uparrow\mathrm{H}1}_\mathrm{epa}
	\pm
	\delta
	\bigl(	
		\cos\theta_K
		\ket{\downarrow\mathrm{H}0}_\mathrm{epa}
		-
		\sin\theta_K
		\ket{\uparrow\mathrm{H}0}_\mathrm{epa}
	\bigr)
	\end{cases}
\end{align}
\end{widetext}
%
with probability
\begin{align}
\begin{aligned}
\nabla_2
	&=
	\left(
		1-
		\nabla
		\cos^2\theta_K
		\sin^2\theta_{N}
	\right)^{N} 
	\\
	&\hspace{0.425cm}
	\left(
		1-
		\nabla
		\sin^2\theta_K
	\right),\label{eq:nabla 2}
\end{aligned}
\end{align}
where $\nabla=\vert\gamma\vert^2\left(\vert\beta\vert^2\right)$. Whenever the photon is found in the transmission channel between Alice and Bob, the electron absorbs it and the protocol declares an erasure.

\item
Alice and Bob repeat the second and third steps for subsequent H(V)-\ac{CEPI}$_N$ gates. After $K$ H(V)-\ac{CEPI}$_N$ gates, unless the photon is absorbed by the electron, 
%
Bob performs the $\M{\mathcal{B}}$ operator again on the output photon to recombine the H and V components of the photon and the composite state of electron-photon pair transforms to:
\begin{align}
\ket{\eta_2}_\text{ep}:
	\begin{cases}
		\ket{\eta^\pm_{20}}_\text{ep}
			=
			\alpha\ket{\downarrow\mathrm{H}}_\text{ep}
			\pm
			\beta\ket{\downarrow\mathrm{V}}_\text{ep},\\
		\ket{\eta^\pm_{21}}_\text{ep}
			=
			\gamma\ket{\uparrow\mathrm{V}}_\text{ep}
			\mp
			\delta\ket{\uparrow\mathrm{H}}_\text{ep}\\
	\end{cases}\label{eq:H(V)-DCFO output}
\end{align}
with probability 
\begin{align} \label{eq:zeta}
\nabla_3
=
	\nabla_2^K.
\end{align}

\end{enumerate}

\begin{figure*}[t!]
 \centering{
\includegraphics[width=0.87\textwidth]{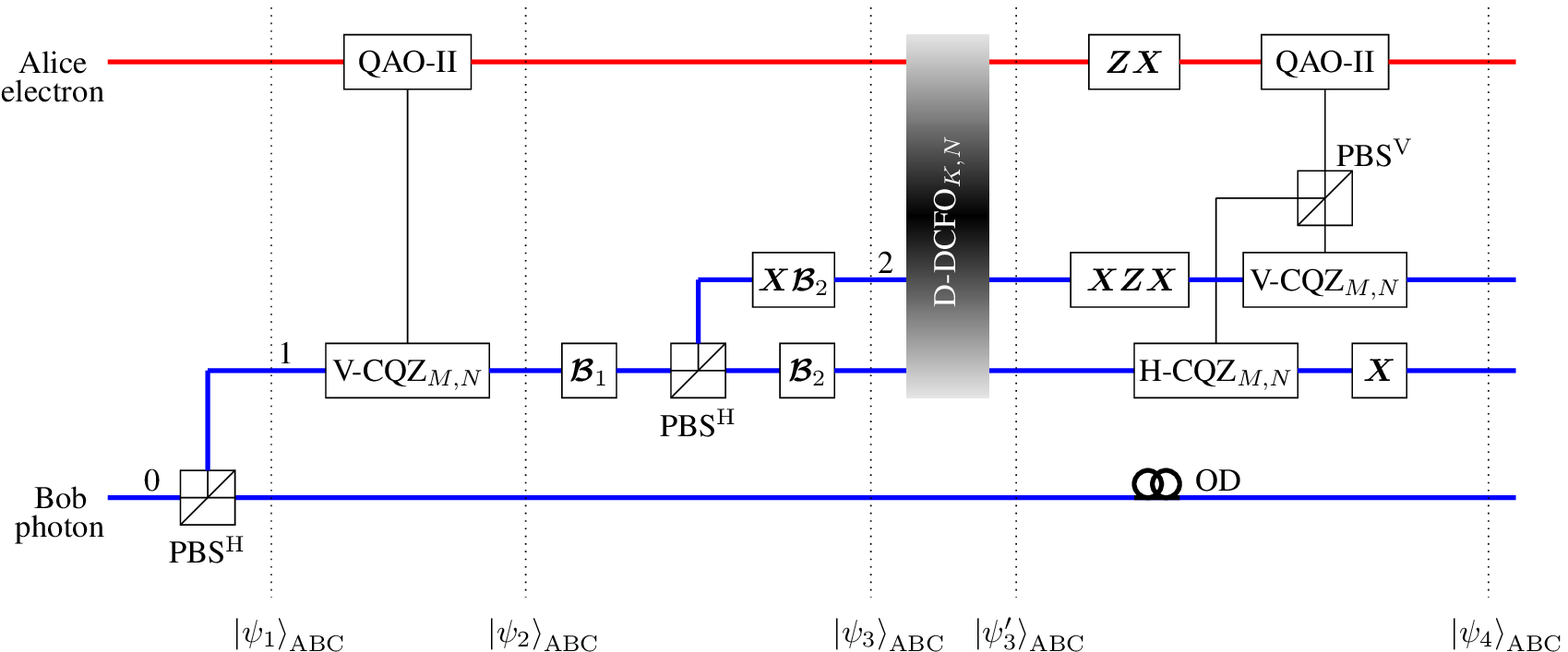}}
    \caption{
A \ac{CCT} protocol for general input states using \ac{QZ} and \ac{CQZ} gates.  Here $\M{\mathcal{B}}_1=\M{R}_\mathrm{z}\hspace{-0.05cm}\left(\varphi\right)\M{X}$ and $\M{\mathcal{B}}_2=\M{R}_\mathrm{z}\hspace{-0.05cm}\left(\phi\right)\M{R}_\mathrm{y}\hspace{-0.05cm}\left(\theta\right)$  where $\M{R}_\mathrm{z}\hspace{-0.05cm}\left(\varphi\right)$ denotes the rotation around $z$-axis. 
}
    \label{fig:QZ-CQZ-BNCU}
\end{figure*}

\subsection{\ac{DDCFO} Gates}\label{sec: 2-4}
Similar to the H(V)-\ac{DCFO}$_{K,N}$ gate, $K$ \ac{DCEPI}$_N$ gates are concatenated serially to devise the \ac{DDCFO} gate where Alice has quantum absorptive object (electron) and Bob equips \ac{QZ}$_N$ gates, as shown in Fig.~\ref{fig:D-DCFO}. To explain the operation of the \ac{DDCFO}$_{K,N}$ gate, consider the initial state of electron-photon pair
\begin{align}
\begin{aligned}
\ket{\eta_0}_\text{epa}
	&=
	\frac{1}{\sqrt{2}}
	\bigl(
		\alpha
		\ket{\downarrow\mathrm{H}}_\mathrm{ep}
		+
		\beta
		\ket{\uparrow\mathrm{V}}_\mathrm{ep}
	\bigr)
	\ket{0}_\mathrm{a}\\
	&\hspace{0.5cm}
	\frac{1}{\sqrt{2}}
	\bigl(
		\gamma
		\ket{\uparrow\mathrm{H}}_\mathrm{ep}
		+
		\delta
		\ket{\downarrow\mathrm{V}}_\mathrm{ep}
	\bigr)
	\ket{1}_\mathrm{a}.
\end{aligned}\label{eq:D-DCFO input}
\end{align}
The \ac{DDCFO}$_{K,N}$ gate takes the following steps. 
\begin{enumerate}
\item 
Bob starts the \ac{DDCFO} by applying PBS$^{\mathrm{V}}$ in each path of the photon and recombines the respective components of the photon after $K$ successive \ac{DCEPI}$_N$ operations. Bob applies the $\M{X}$ operator on the photon component in path state $\ket{0}_\mathrm{a}$.

\item
Now, Alice performs $\M{R}_\mathrm{y}\hspace{-0.05cm}\left(2\theta_K\right)$ on her qubit and Bob inputs the photon components in path states $\ket{0}_\mathrm{a}$ and $\ket{1}_\mathrm{a}$ to the \ac{DCEPI}$_N$ gate. Unless the photon is absorbed by the electron, the first \ac{DCEPI}$_N$ gate transforms the $\ket{\eta_0}_\text{epa}$ as
\begin{align}
\begin{aligned}
\ket{\eta_1}_\text{epa}
	&=
	\frac{1}{\sqrt{2}}
	\big(
		\beta
		\ket{\uparrow\mathrm{H}0}_\mathrm{epa}
		+
		\delta
		\ket{\downarrow\mathrm{V}1}_\mathrm{epa}
		\big.
		\\
		&\big.\hspace{0.55cm}
		+~\alpha	
		\cos\theta_K
		\ket{\downarrow\mathrm{H}2}_\mathrm{epa}
		\big.
		\\
		&\big.\hspace{0.55cm}
		-~
		\alpha
		\sin\theta_K
		\ket{\uparrow\mathrm{H}2}_\mathrm{epa}
		\big.
		\\
		&\big.\hspace{0.55cm}
		+~
		\gamma	
		\cos\theta_K
		\ket{\uparrow\mathrm{H}3}_\mathrm{epa}
		\big.
		\\
		&\big.\hspace{0.55cm}
		+~
		\gamma
		\sin\theta_K
		\ket{\downarrow\mathrm{H}3}_\mathrm{epa}
	\big)
\end{aligned}
\end{align}
%
with probability
\begin{align}
\begin{aligned}
\nabla_5
	&=
	\left(
		1-
		\nabla_4
		\cos^2\theta_K
		\sin^2\theta_{N}
	\right)^{N} 
	\\
	&\hspace{0.425cm}
	\left(
		1-
		\nabla_4
		\sin^2\theta_K
	\right),
\end{aligned}
\end{align}
where $\nabla_4=\vert\beta\vert^2+\vert\delta\vert^2$. 

\item 
Alice and Bob repeat the second step for subsequent \ac{DCEPI}$_N$ gates. After $K$ \ac{DCEPI}$_N$ gates, unless the photon is absorbed by the electron, 
%
Bob performs the $\M{X}$ operator on the photon component in path state $\ket{0}_\mathrm{a}$ and recombines the respective photon components. Then, the composite state of electron-photon pair transforms to
\begin{align}
\begin{aligned}
\ket{\eta_2}_\text{epa}
	=&
	\frac{1}{\sqrt{2}}
	\bigl(
		-\alpha
		\ket{\uparrow\mathrm{H}}_\mathrm{ep}
		+
		\beta
		\ket{\uparrow\mathrm{V}}_\mathrm{ep}
	\bigr)
	\ket{0}_\mathrm{a}\\
	&
	\frac{1}{\sqrt{2}}
	\bigl(
		\gamma
		\ket{\downarrow\mathrm{H}}_\mathrm{ep}
		+
		\delta
		\ket{\downarrow\mathrm{V}}_\mathrm{ep}
	\bigr)
	\ket{1}_\mathrm{a}
\end{aligned}\label{eq:D-DCFO output}
\end{align}
with probability
\begin{align}
\nabla_6
=
	\nabla_5^K.
\end{align}

\end{enumerate}

\subsection{\ac{CCT} for General Input States}\label{sec: 2-5}
For the \ac{CCT} with general input states, Alice and Bob prepare the target and control qubits in electron and photon: $\ket{\psi}_\text{A}=\ket{\text{electron}}_\mathrm{e}$ and  $\ket{\psi}_\text{B}=\ket{\text{photon}}_\mathrm{p}$ where
\begin{align}
\begin{aligned} 
\ket{0}_{\mathrm{A}}
	&=
	\ket{\uparrow}_{\mathrm{e}}, 
	\\
\ket{1}_{\mathrm{A}}
	&=
	\ket{\downarrow}_{\mathrm{e}}, \\
\ket{0}_{\mathrm{B}}
	&=
	\ket{\mathrm{H}}_{\mathrm{p}},\\ 
\ket{1}_{\mathrm{B}}
	&=
	\ket{\mathrm{V}}_{\mathrm{p}}.
	\label{eq:B:1}
\end{aligned}
\end{align}
To devise the \ac{CCT}, Alice and Bob takes the following steps (see Fig.~\ref{fig:QZ-CQZ-BNCU}).
\begin{enumerate}
\item
Bob starts the protocol by throwing his photon towards PBS$^\mathrm{H}$ to entangle the polarization (control qubit) state $\ket{\psi}_\mathrm{B}$ with path state $\ket{0}_\mathrm{C}$. Then Alice and Bob have the composite state $\ket{\psi_1}_\mathrm{ABC}$ in \eqref{eq:psi 1}.

\item 
Bob detours the H component of the photon to recombine it at the end of the protocol and inputs the V component of the photon to the V-\ac{CQZ}$_{M,N}$ gate. Unless the photon is absorbed by the electron or discarded at the detector in V-\ac{CQZ}$_{M,N}$ gate, it transforms $\ket{\psi_1}_\mathrm{ABC}$ to $\ket{\psi_2}_\mathrm{ABC}$ in \eqref{eq:psi 2} with probability 
\begin{align}
\begin{aligned}
\lambda_2
	&=
	\left(
		1-
		\vert
			\alpha
			\delta
		\vert^2
		\sin^2\theta_M
	\right)^M
	\\
	&\hspace{0.425cm}
	\prod_{i=1}^M	
	\left[
		1-
		\vert
			\beta
			\delta
		\vert^2
		\sin^2\left(i\theta_M\right)
		\sin^2\theta_N
	\right]^N
\end{aligned}
\end{align}
tending to one as $M,N\rightarrow\infty$. 

\item
Bob applies the $\M{\mathcal{B}}_1=\M{R}_\mathrm{z}\hspace{-0.05cm}\left(\varphi\right)\M{X}$ operation followed by PBS$^\mathrm{H}$ on the component of the photon in path state $\ket{1}_\mathrm{C}$.
\item 
Bob applies $\M{\mathcal{B}}_2=\M{R}_\mathrm{z}\hspace{-0.05cm}\left(\phi\right)\M{R}_\mathrm{y}\hspace{-0.05cm}\left(\theta\right)$ and $\M{X}\M{\mathcal{B}}_2$ operators on the photon components in the path states $\ket{1}_\mathrm{C}$ and $\ket{2}_\mathrm{C}$, respectively.
Then, the composite state of Alice and Bob transforms to $\ket{\psi_3}_\text{ABC}$ in \eqref{eq:psi 3}.

\item 
Bob inputs the photon components in path states $\ket{1}_\mathrm{C}$ and $\ket{2}_\mathrm{C}$ to the D-DCFO$_{K,N}$ gate. From \eqref{eq:D-DCFO input}--\eqref{eq:D-DCFO output}, unless the photon is absorbed by the electron in the \ac{DDCFO}$_{K,N}$ gate, it transforms $\ket{\psi_3}_\mathrm{ABC}$ as follows: 
\begin{equation}
\begin{aligned}
\ket{\psi'_3}_{\text{ABC}}
	&=
	\gamma
	\ket{\psi}_\mathrm{A}
	\ket{00}_\text{BC}
	\\
	&\hspace{0.35cm}+
	\delta
	\alpha 
	e^{-\iota\left(\varphi+\phi\right)/2}
	\cos\left(\theta/2\right)
	\ket{101}_{\text{ABC}}
	\\
	&\hspace{0.35cm}+
	\delta
	\alpha 
	e^{-\iota\left(\varphi-\phi\right)/2}
	\sin\left(\theta/2\right)
	\ket{011}_{\text{ABC}}
	\\
	&\hspace{0.35cm}-
	\delta
	\beta 
	e^{\iota\left(\varphi+\phi\right)/2}
	\cos\left(\theta/2\right)
	\ket{002}_{\text{ABC}}
	\\
	&\hspace{0.35cm}-
	\delta
	\beta 
	e^{\iota\left(\varphi-\phi\right)/2}
	\sin\left(\theta/2\right)
	\ket{112}_{\text{ABC}}
\end{aligned}
\end{equation}
%
with probability
\begin{align}
\begin{aligned}
\lambda_3
	&=
	\left(
		1-
		\vert\delta\vert^2
		\sin^2\left(\theta/2\right)
		\cos^2\theta_K
		\sin^2\theta_N
	\right)^{KN}
	\\
	&\hspace{0.425cm}
	\left(
		1-
		\vert\delta\vert^2
		\sin^2\left(\theta/2\right)
		\sin^2\theta_K
	\right)^K.
\end{aligned}
\end{align}

\item 
Alice applies the $\M{X}$ followed by the $\M{Z}$ on her qubit and Bob performs $\M{X}\M{Z}\M{X}$ operation on the path state    $\ket{2}_\mathrm{C}$, respectively. 

\item 
Bob inputs the component of the photon in path state $\ket{1}_\mathrm{C}$ to the H-\ac{CQZ}$_{M,N}$ gate and $\ket{2}_\mathrm{C}$ to the V-\ac{CQZ}$_{M,N}$ gate, respectively. Unless the photon is discarded in the \ac{CQZ}$_{M,N}$ gates, Bob applies $\M{X}$ operator on the component of photon in path state $\ket{1}_\mathrm{C}$ and the composite of Alice and Bob transforms to $\ket{\psi_4}_\mathrm{ABC}$ in \eqref{eq:psi 4} \footnote{In the presence of the absorptive object, the H(V)-\ac{CQZ}$_{M,N}$ gate transforms $\ket{\mathrm{V}\left(\mathrm{H}\right)}_{\mathrm{p}}$ to $-\ket{\mathrm{H}\left(\mathrm{V}\right)}_{\mathrm{p}}$.}
with probability
\begin{align}
\begin{aligned}
\lambda_4
	&=
	\left(
		1-
		\nabla_7
		\sin^2\theta_M
	\right)^M
	\\
	&\hspace{0.425cm}
	\prod_{i=1}^M	
	\left[
		1-
		\nabla_8
		\sin^2\left(i\theta_M\right)
		\sin^2\theta_N
	\right]^N,
\end{aligned}
\end{align} 
where $\nabla_7$ and $\nabla_8$ are defined as:
\begin{align}
\nabla_7
	&=
	\vert
		\delta
		\alpha
	\vert^2
	\cos^2\left(\theta/2\right)
	+
	\vert
		\delta
		\beta
	\vert^2
	\sin^2\left(\theta/2\right),
	\\
\nabla_8
	&=
	\vert
		\delta
		\beta
	\vert^2
	\cos^2\left(\theta/2\right)
	+
	\vert
		\delta
		\alpha
	\vert^2
	\sin^2\left(\theta/2\right).
\end{align}

\item
To disentangle the ancillary qutrit, Bob applies the local operation $\M{V}_2$ followed by the Hadamard gate $\M{H}$ on the ancilla and performs the measurement on the ancilla in the computational basis. It collapses the composite state $\ket{\psi_4}_{\text{ABC}}$ to $\ket{\psi_{5m}}_\mathrm{AB}$ in \eqref{eq:psi 5}.

\begin{figure}[t!]
 \centering{
\includegraphics[width=0.35\textwidth]{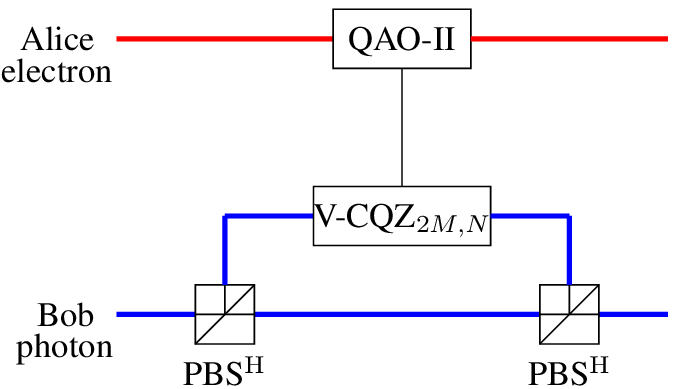}}
    \caption{
   A counterfactual controlled-$\M{Z}$ operator.    
   }
    \label{fig:CCZ}
\end{figure}

\item
Alice and Bob apply the global unitary operation $\M{Q}_3$ depending on
Bob's announcement of the measurement result $m$ with classical communication. For $m=0$, $\M{Q}_3=\M{I}$. For $m=1$, Alice applies the $\M{Z}$ operator on her qubit and Bob applies $\M{X}$. Bob throws his photon towards PBS$^{\mathrm{H}}$ again and inputs the V components of the photon to the V-\ac{CQZ}$_{2M,N}$ gate with $2M$ outer cycles to apply the controlled-$\M{Z}$ operator counterfactually as shown in Fig.~\ref{fig:CCZ}.  
Unless the photon is discarded in the \ac{CQZ} gate, 
%
Bob recombines the H and V components of the photon with probability
\begin{align}
\begin{aligned}
\lambda_5
	&=
	\left(
		1-
		\vert
			\alpha
			\gamma
		\vert^2
		\sin^2\theta_M
	\right)^{2M}
	\\
	&\hspace{0.35cm}
	\prod_{i=1}^{2M}	
	\left[
		1-
		\vert
			\beta
			\gamma
		\vert^2
		\sin^2\left(i\theta_M\right)
		\sin^2\theta_N
	\right]^N,
\end{aligned}
\end{align}
followed by applying the $\M{X}$ operator on his qubit.
Unless the photon is discarded, the composite state $\ket{\psi_{5m}}_\mathrm{AB}$ transforms to $\ket{\psi_{6m}}_\mathrm{AB}$ in \eqref{eq:psi 6} with probability $\lambda_5^m$.

\end{enumerate}
In the CCT protocol for general input states, there exists the nonzero probability (abortion rate) $\zeta_m$ that the photon is traveled over the quantum channel and the protocol fails in counterfactuality where 
\begin{align}
\label{eq:FP}
\zeta_m
	=
	1-
	\lambda_2
	\lambda_3
	\lambda_4
	\lambda_5^m,
\end{align}
which tends to zero as $M,N,K\rightarrow\infty$.

\begin{figure*}[t!]
 \centering{
\includegraphics[width=0.63\textwidth]{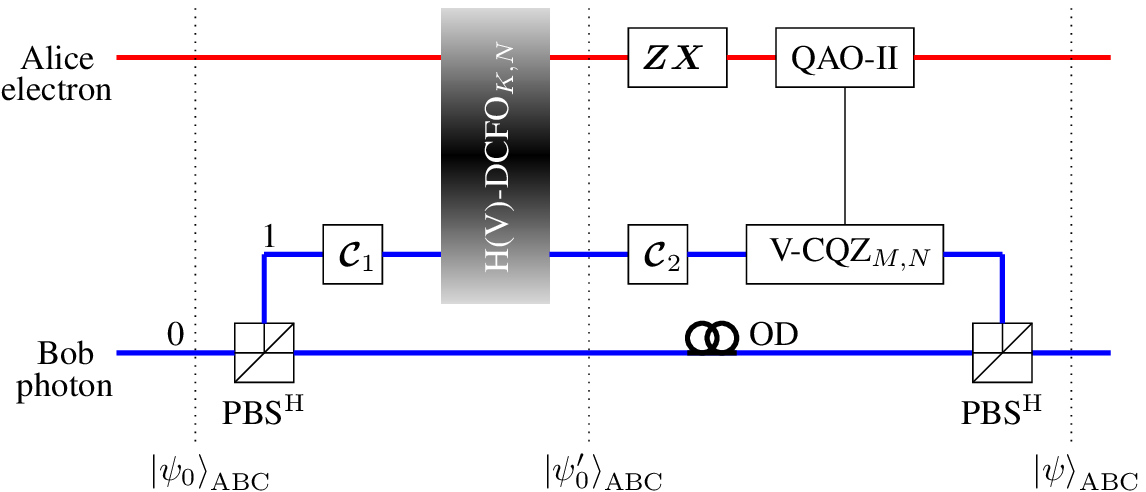}}
    \caption{
A \ac{CCT} protocol using the H(V) DCFO$_{K,N}$ gate for Bell-type states of 1(0)-class. Here $\M{\mathcal{C}}_1=\M{X}^{1-\ell}\M{U}\M{X}^\ell$, $\M{\mathcal{C}}_2=\M{X}\M{Z}^{1-\ell}\M{X}^{1-\ell}$ for $\ell$-class states and $\ket{\psi_0}_\mathrm{ABC}=\ket{\psi_0}_\mathrm{AB}\otimes\ket{0}_\mathrm{C}$.
}
    \label{fig:QZ-CQZ-BTS-BNCU}
\end{figure*}

\subsection{\ac{CCT} for Bell-Type Input States}\label{sec: 2-6}
To demonstrate the implementation of \ac{CCT} for Bell-type states, consider that the composite input state of Alice and Bob is $\ket{\psi_0}_\mathrm{AB}=\ket{\psi^\pm_{0\ell}}_\text{AB}$,  $\ell=0,1$, in (19) of the main text. 
Similar to the general setup, Bob starts the protocol by entangling his qubit with the ancillary qubit $\ket{0}_\mathrm{C}$ by throwing his photon towards PBS$^\mathrm{H}$ as shown in Fig.~\ref{fig:QZ-CQZ-BTS-BNCU}. The \ac{CCT} protocol for Bell-type states of 1(0)-class takes the following steps.

\begin{enumerate}
\item 
Bob applies $\M{\mathcal{C}}_1$ on the component of the photon in path state $\ket{1}_\mathrm{C}$ where $\M{\mathcal{C}}_1=\M{X}^{1-\ell}\M{U}\M{X}^\ell$ for $\ell$-class states.

\item 
Bob inputs the photon component in path state $\ket{1}_\mathrm{C}$ to the H(V)-\ac{DCFO}$_{K,N}$ gate. From \eqref{eq:H(V)-DCFO input}--\eqref{eq:H(V)-DCFO output}, unless the photon is absorbed by the electron, the H(V)-\ac{DCFO}$_{K,N}$ gate transforms the composite of Alice and Bob as follows: 
\begin{widetext}
\begin{align}
\ket{\psi'_{0}}_\text{ABC}:
	\begin{cases}
	\ket{\psi'^{\pm}_{00}}_\text{ABC}
	=
	\alpha
	\ket{100}_\mathrm{ABC}
	\pm
	\beta
	\left(
			-e^{\iota\left(\varphi+\phi\right)/2}
			\cos\left(\theta/2\right)
			\ket{001}_\mathrm{ABC}
			-
			e^{\iota\left(\varphi-\phi\right)/2}
			\sin\left(\theta/2\right)
			\ket{111}_\mathrm{ABC}
	\right),
	\\
	\ket{\psi'^{\pm}_{01}}_\text{ABC}
	=
	\gamma
	\left(
			e^{\iota\left(\varphi+\phi\right)/2}
			\cos\left(\theta/2\right)
			\ket{101}_\mathrm{ABC}
			+
			e^{\iota\left(\varphi-\phi\right)/2}
			\sin\left(\theta/2\right)
			\ket{011}_\mathrm{ABC}
	\right)
	\mp
	\delta
	\ket{000}_\mathrm{ABC}
	\end{cases}
\end{align}
\end{widetext}
%
with probability
\begin{align}
\begin{aligned}
\lambda_6
	&=
	\left(
		1-
		\nabla
		\sin^2\left(\theta/2\right)
		\cos^2\theta_K
		\sin^2\theta_N
	\right)^{KN}
	\\
	&\hspace{0.425cm}
	\left(
		1-
		\nabla
		\sin^2\left(\theta/2\right)
		\sin^2\theta_K
	\right)^K.
\end{aligned}
\end{align}

\item 
Alice applies the $\M{X}$ followed by the $\M{Z}$ on her qubit and Bob performs $\M{\mathcal{C}}_2=\M{X}\M{Z}^{1-\ell}\M{X}^{1-\ell}$ for $\ell$-class states on the photon component in path state $\ket{1}_\mathrm{C}$, respectively. 
\item 
Bob inputs the photon component in path state $\ket{1}_\mathrm{C}$ to the V-\ac{CQZ}$_{M,N}$ gate and recombines the photon component in path state $\ket{0}_\mathrm{C}$ and $\ket{1}_\mathrm{C}$ after the V-\ac{CQZ}$_{M,N}$ gate. At the end of the protocol, unless the photon is discarded in the V-\ac{CQZ}$_{M,N}$ gate,  Bob recombines the H and V components of the photon and the composite state $\ket{\psi'_{0}}_\text{ABC}$ transforms to $\ket{\psi}_\mathrm{ABC}$ in (23) of the main text with probability 
\begin{align}
\begin{aligned}
\lambda_7
	&=
	\left(
		1-
		\nabla_{9\left(10\right)}
		\sin^2\theta_M
	\right)^M
	\\
	&\hspace{0.35cm}
	\prod_{i=1}^{M}	
	\left[
		1-
		\nabla_{10\left(9\right)}
		\sin^2\left(i\theta_M\right)
		\sin^2\theta_N
	\right]^N
\end{aligned}
\end{align}
for $1\left(0\right)$-class states where 
\begin{align}
\nabla_{9}
&=\nabla\cos^{2}\left(\theta/2\right),
\\
\nabla_{10}
&=\nabla\sin^{2}\left(\theta/2\right).
\end{align}
%

\end{enumerate}
For the CCT protocol with Bell-type states, the probability (abortion rate) that the protocol fails to ensure the counterfactuality is given by  
\begin{align}
\zeta
	=
	1
	-
	\lambda_6
	\lambda_7
\end{align}
tending again to zero as $M,N,K\rightarrow\infty$.

This work was supported in part by the the National Research Foundation of Korea under Grant  2019R1A2C2007037 and in part by the Office of Naval Research under Grant  N00014-19-1-2724.


\begin{thebibliography}{31}%
\makeatletter
\providecommand \@ifxundefined [1]{%
 \@ifx{#1\undefined}
}%
\providecommand \@ifnum [1]{%
 \ifnum #1\expandafter \@firstoftwo
 \else \expandafter \@secondoftwo
 \fi
}%
\providecommand \@ifx [1]{%
 \ifx #1\expandafter \@firstoftwo
 \else \expandafter \@secondoftwo
 \fi
}%
\providecommand \natexlab [1]{#1}%
\providecommand \enquote  [1]{``#1''}%
\providecommand \bibnamefont  [1]{#1}%
\providecommand \bibfnamefont [1]{#1}%
\providecommand \citenamefont [1]{#1}%
\providecommand \href@noop [0]{\@secondoftwo}%
\providecommand \href [0]{\begingroup \@sanitize@url \@href}%
\providecommand \@href[1]{\@@startlink{#1}\@@href}%
\providecommand \@@href[1]{\endgroup#1\@@endlink}%
\providecommand \@sanitize@url [0]{\catcode `\\12\catcode `\$12\catcode
  `\&12\catcode `\#12\catcode `\^12\catcode `\_12\catcode `\%12\relax}%
\providecommand \@@startlink[1]{}%
\providecommand \@@endlink[0]{}%
\providecommand \url  [0]{\begingroup\@sanitize@url \@url }%
\providecommand \@url [1]{\endgroup\@href {#1}{\urlprefix }}%
\providecommand \urlprefix  [0]{URL }%
\providecommand \Eprint [0]{\href }%
\providecommand \doibase [0]{https://doi.org/}%
\providecommand \selectlanguage [0]{\@gobble}%
\providecommand \bibinfo  [0]{\@secondoftwo}%
\providecommand \bibfield  [0]{\@secondoftwo}%
\providecommand \translation [1]{[#1]}%
\providecommand \BibitemOpen [0]{}%
\providecommand \bibitemStop [0]{}%
\providecommand \bibitemNoStop [0]{.\EOS\space}%
\providecommand \EOS [0]{\spacefactor3000\relax}%
\providecommand \BibitemShut  [1]{\csname bibitem#1\endcsname}%
\let\auto@bib@innerbib\@empty
\bibitem [{\citenamefont {Cirac}\ \emph {et~al.}(1999)\citenamefont {Cirac},
  \citenamefont {Ekert}, \citenamefont {Huelga},\ and\ \citenamefont
  {Macchiavello}}]{CEHM:99:PRA}%
  \BibitemOpen
  \bibfield  {author} {\bibinfo {author} {\bibfnamefont {J.~I.}\ \bibnamefont
  {Cirac}}, \bibinfo {author} {\bibfnamefont {A.~K.}\ \bibnamefont {Ekert}},
  \bibinfo {author} {\bibfnamefont {S.~F.}\ \bibnamefont {Huelga}},\ and\
  \bibinfo {author} {\bibfnamefont {C.}~\bibnamefont {Macchiavello}},\
  }\href@noop {} {\bibfield  {journal} {\bibinfo  {journal} {Phys. Rev. A}\
  }\textbf {\bibinfo {volume} {59}},\ \bibinfo {pages} {4249} (\bibinfo {year}
  {1999})}\BibitemShut {NoStop}%
\bibitem [{\citenamefont {Dai}\ \emph {et~al.}(2020)\citenamefont {Dai},
  \citenamefont {Peng},\ and\ \citenamefont {Win}}]{DaiPenWin:J20b}%
  \BibitemOpen
  \bibfield  {author} {\bibinfo {author} {\bibfnamefont {W.}~\bibnamefont
  {Dai}}, \bibinfo {author} {\bibfnamefont {T.}~\bibnamefont {Peng}},\ and\
  \bibinfo {author} {\bibfnamefont {M.~Z.}\ \bibnamefont {Win}},\ }\href@noop
  {} {\bibfield  {journal} {\bibinfo  {journal} {{IEEE} J. Sel. Areas Commun.}\
  }\textbf {\bibinfo {volume} {38}},\ \bibinfo {pages} {540} (\bibinfo {year}
  {2020})},\ \bibinfo {note} {special issue on {\em Advances in Quantum
  Communications, Computing, Cryptography and Sensing}}\BibitemShut {NoStop}%
\bibitem [{\citenamefont {Cacciapuoti}\ \emph {et~al.}(2019)\citenamefont
  {Cacciapuoti}, \citenamefont {Caleffi}, \citenamefont {Tafuri}, \citenamefont
  {Cataliotti}, \citenamefont {Gherardini},\ and\ \citenamefont
  {Bianchi}}]{CCTCGB:19:IEEEN}%
  \BibitemOpen
  \bibfield  {author} {\bibinfo {author} {\bibfnamefont {A.~S.}\ \bibnamefont
  {Cacciapuoti}}, \bibinfo {author} {\bibfnamefont {M.}~\bibnamefont
  {Caleffi}}, \bibinfo {author} {\bibfnamefont {F.}~\bibnamefont {Tafuri}},
  \bibinfo {author} {\bibfnamefont {F.~S.}\ \bibnamefont {Cataliotti}},
  \bibinfo {author} {\bibfnamefont {S.}~\bibnamefont {Gherardini}},\ and\
  \bibinfo {author} {\bibfnamefont {G.}~\bibnamefont {Bianchi}},\ }\href@noop
  {} {\bibfield  {journal} {\bibinfo  {journal} {IEEE Network}\ }\textbf
  {\bibinfo {volume} {34}},\ \bibinfo {pages} {137} (\bibinfo {year}
  {2019})}\BibitemShut {NoStop}%
\bibitem [{\citenamefont {Eisert}\ \emph {et~al.}(2000)\citenamefont {Eisert},
  \citenamefont {Jacobs}, \citenamefont {Papadopoulos},\ and\ \citenamefont
  {Plenio}}]{EJPP:00:PRA}%
  \BibitemOpen
  \bibfield  {author} {\bibinfo {author} {\bibfnamefont {J.}~\bibnamefont
  {Eisert}}, \bibinfo {author} {\bibfnamefont {K.}~\bibnamefont {Jacobs}},
  \bibinfo {author} {\bibfnamefont {P.}~\bibnamefont {Papadopoulos}},\ and\
  \bibinfo {author} {\bibfnamefont {M.~B.}\ \bibnamefont {Plenio}},\
  }\href@noop {} {\bibfield  {journal} {\bibinfo  {journal} {Phys. Rev. A}\
  }\textbf {\bibinfo {volume} {62}},\ \bibinfo {pages} {052317} (\bibinfo
  {year} {2000})}\BibitemShut {NoStop}%
\bibitem [{\citenamefont {Fletcher}\ \emph {et~al.}(2008)\citenamefont
  {Fletcher}, \citenamefont {Shor},\ and\ \citenamefont
  {Win}}]{FleShoWin:J08a}%
  \BibitemOpen
  \bibfield  {author} {\bibinfo {author} {\bibfnamefont {A.~S.}\ \bibnamefont
  {Fletcher}}, \bibinfo {author} {\bibfnamefont {P.~W.}\ \bibnamefont {Shor}},\
  and\ \bibinfo {author} {\bibfnamefont {M.~Z.}\ \bibnamefont {Win}},\
  }\href@noop {} {\bibfield  {journal} {\bibinfo  {journal} {{IEEE} Trans. Inf.
  Theory}\ }\textbf {\bibinfo {volume} {54}},\ \bibinfo {pages} {5705}
  (\bibinfo {year} {2008})}\BibitemShut {NoStop}%
\bibitem [{\citenamefont {Chiani}\ \emph {et~al.}(2020)\citenamefont {Chiani},
  \citenamefont {Conti},\ and\ \citenamefont {Win}}]{ChiConWin:J20}%
  \BibitemOpen
  \bibfield  {author} {\bibinfo {author} {\bibfnamefont {M.}~\bibnamefont
  {Chiani}}, \bibinfo {author} {\bibfnamefont {A.}~\bibnamefont {Conti}},\ and\
  \bibinfo {author} {\bibfnamefont {M.~Z.}\ \bibnamefont {Win}},\ }\href
  {https://doi.org/10.1103/PhysRevA.102.012410} {\bibfield  {journal} {\bibinfo
   {journal} {Phys. Rev. A}\ }\textbf {\bibinfo {volume} {102}},\ \bibinfo
  {pages} {012410} (\bibinfo {year} {2020})}\BibitemShut {NoStop}%
\bibitem [{\citenamefont {Chen}\ and\ \citenamefont {Yu}(2015)}]{CY:15:PRA}%
  \BibitemOpen
  \bibfield  {author} {\bibinfo {author} {\bibfnamefont {L.}~\bibnamefont
  {Chen}}\ and\ \bibinfo {author} {\bibfnamefont {L.}~\bibnamefont {Yu}},\
  }\href@noop {} {\bibfield  {journal} {\bibinfo  {journal} {Phys. Rev. A}\
  }\textbf {\bibinfo {volume} {91}},\ \bibinfo {pages} {032308} (\bibinfo
  {year} {2015})}\BibitemShut {NoStop}%
\bibitem [{\citenamefont {Barenco}\ \emph {et~al.}(1995)\citenamefont
  {Barenco}, \citenamefont {Bennett}, \citenamefont {Cleve}, \citenamefont
  {DiVincenzo}, \citenamefont {Margolus}, \citenamefont {Shor}, \citenamefont
  {Sleator}, \citenamefont {Smolin},\ and\ \citenamefont
  {Weinfurter}}]{BBCDMSSSW:95:PRA}%
  \BibitemOpen
  \bibfield  {author} {\bibinfo {author} {\bibfnamefont {A.}~\bibnamefont
  {Barenco}}, \bibinfo {author} {\bibfnamefont {C.~H.}\ \bibnamefont
  {Bennett}}, \bibinfo {author} {\bibfnamefont {R.}~\bibnamefont {Cleve}},
  \bibinfo {author} {\bibfnamefont {D.~P.}\ \bibnamefont {DiVincenzo}},
  \bibinfo {author} {\bibfnamefont {N.}~\bibnamefont {Margolus}}, \bibinfo
  {author} {\bibfnamefont {P.}~\bibnamefont {Shor}}, \bibinfo {author}
  {\bibfnamefont {T.}~\bibnamefont {Sleator}}, \bibinfo {author} {\bibfnamefont
  {J.~A.}\ \bibnamefont {Smolin}},\ and\ \bibinfo {author} {\bibfnamefont
  {H.}~\bibnamefont {Weinfurter}},\ }\href@noop {} {\bibfield  {journal}
  {\bibinfo  {journal} {Phys. Rev. A}\ }\textbf {\bibinfo {volume} {52}},\
  \bibinfo {pages} {3457} (\bibinfo {year} {1995})}\BibitemShut {NoStop}%
\bibitem [{\citenamefont {Chen}\ and\ \citenamefont {Yu}(2014)}]{CY:14:PRA}%
  \BibitemOpen
  \bibfield  {author} {\bibinfo {author} {\bibfnamefont {L.}~\bibnamefont
  {Chen}}\ and\ \bibinfo {author} {\bibfnamefont {L.}~\bibnamefont {Yu}},\
  }\href@noop {} {\bibfield  {journal} {\bibinfo  {journal} {Phys. Rev. A}\
  }\textbf {\bibinfo {volume} {89}},\ \bibinfo {pages} {062326} (\bibinfo
  {year} {2014})}\BibitemShut {NoStop}%
\bibitem [{\citenamefont {Soeda}\ \emph {et~al.}(2011)\citenamefont {Soeda},
  \citenamefont {Turner},\ and\ \citenamefont {Murao}}]{STM:11:PRL}%
  \BibitemOpen
  \bibfield  {author} {\bibinfo {author} {\bibfnamefont {A.}~\bibnamefont
  {Soeda}}, \bibinfo {author} {\bibfnamefont {P.~S.}\ \bibnamefont {Turner}},\
  and\ \bibinfo {author} {\bibfnamefont {M.}~\bibnamefont {Murao}},\
  }\href@noop {} {\bibfield  {journal} {\bibinfo  {journal} {Phys. Rev. Lett.}\
  }\textbf {\bibinfo {volume} {107}},\ \bibinfo {pages} {180501} (\bibinfo
  {year} {2011})}\BibitemShut {NoStop}%
\bibitem [{\citenamefont {Lemr}\ \emph {et~al.}(2015)\citenamefont {Lemr},
  \citenamefont {Bartkiewicz}, \citenamefont {{\v{C}}ernoch}, \citenamefont
  {Du{\v{s}}ek},\ and\ \citenamefont {Soubusta}}]{LBCDS:15:PRL}%
  \BibitemOpen
  \bibfield  {author} {\bibinfo {author} {\bibfnamefont {K.}~\bibnamefont
  {Lemr}}, \bibinfo {author} {\bibfnamefont {K.}~\bibnamefont {Bartkiewicz}},
  \bibinfo {author} {\bibfnamefont {A.}~\bibnamefont {{\v{C}}ernoch}}, \bibinfo
  {author} {\bibfnamefont {M.}~\bibnamefont {Du{\v{s}}ek}},\ and\ \bibinfo
  {author} {\bibfnamefont {J.}~\bibnamefont {Soubusta}},\ }\href@noop {}
  {\bibfield  {journal} {\bibinfo  {journal} {Phys. Rev. Lett.}\ }\textbf
  {\bibinfo {volume} {114}},\ \bibinfo {pages} {153602} (\bibinfo {year}
  {2015})}\BibitemShut {NoStop}%
\bibitem [{\citenamefont {Giovannetti}\ \emph {et~al.}(2013)\citenamefont
  {Giovannetti}, \citenamefont {Maccone}, \citenamefont {Morimae},\ and\
  \citenamefont {Rudolph}}]{GMMR:13:PRL}%
  \BibitemOpen
  \bibfield  {author} {\bibinfo {author} {\bibfnamefont {V.}~\bibnamefont
  {Giovannetti}}, \bibinfo {author} {\bibfnamefont {L.}~\bibnamefont
  {Maccone}}, \bibinfo {author} {\bibfnamefont {T.}~\bibnamefont {Morimae}},\
  and\ \bibinfo {author} {\bibfnamefont {T.~G.}\ \bibnamefont {Rudolph}},\
  }\href@noop {} {\bibfield  {journal} {\bibinfo  {journal} {Phys. Rev. Lett.}\
  }\textbf {\bibinfo {volume} {111}},\ \bibinfo {pages} {230501} (\bibinfo
  {year} {2013})}\BibitemShut {NoStop}%
\bibitem [{\citenamefont {P\'erez-Delgado}\ and\ \citenamefont
  {Fitzsimons}(2015)}]{PF:15:PRL}%
  \BibitemOpen
  \bibfield  {author} {\bibinfo {author} {\bibfnamefont {C.~A.}\ \bibnamefont
  {P\'erez-Delgado}}\ and\ \bibinfo {author} {\bibfnamefont {J.~F.}\
  \bibnamefont {Fitzsimons}},\ }\href@noop {} {\bibfield  {journal} {\bibinfo
  {journal} {Phys. Rev. Lett.}\ }\textbf {\bibinfo {volume} {114}},\ \bibinfo
  {pages} {220502} (\bibinfo {year} {2015})}\BibitemShut {NoStop}%
\bibitem [{\citenamefont {Fitzsimons}(2017)}]{F:17:NPJQI}%
  \BibitemOpen
  \bibfield  {author} {\bibinfo {author} {\bibfnamefont {J.~F.}\ \bibnamefont
  {Fitzsimons}},\ }\href@noop {} {\bibfield  {journal} {\bibinfo  {journal}
  {npj Quantum Inf.}\ }\textbf {\bibinfo {volume} {3}},\ \bibinfo {pages} {1}
  (\bibinfo {year} {2017})}\BibitemShut {NoStop}%
\bibitem [{\citenamefont {Salih}\ \emph {et~al.}(2013)\citenamefont {Salih},
  \citenamefont {Li}, \citenamefont {Al-Amri},\ and\ \citenamefont
  {Zubairy}}]{SLAZ:13:PRL}%
  \BibitemOpen
  \bibfield  {author} {\bibinfo {author} {\bibfnamefont {H.}~\bibnamefont
  {Salih}}, \bibinfo {author} {\bibfnamefont {Z.-H.}\ \bibnamefont {Li}},
  \bibinfo {author} {\bibfnamefont {M.}~\bibnamefont {Al-Amri}},\ and\ \bibinfo
  {author} {\bibfnamefont {M.~S.}\ \bibnamefont {Zubairy}},\ }\href@noop {}
  {\bibfield  {journal} {\bibinfo  {journal} {Phys. Rev. Lett.}\ }\textbf
  {\bibinfo {volume} {110}},\ \bibinfo {pages} {170502} (\bibinfo {year}
  {2013})}\BibitemShut {NoStop}%
\bibitem [{\citenamefont {Aharonov}\ and\ \citenamefont
  {Vaidman}(2019)}]{AV:19:PRA}%
  \BibitemOpen
  \bibfield  {author} {\bibinfo {author} {\bibfnamefont {Y.}~\bibnamefont
  {Aharonov}}\ and\ \bibinfo {author} {\bibfnamefont {L.}~\bibnamefont
  {Vaidman}},\ }\href@noop {} {\bibfield  {journal} {\bibinfo  {journal} {Phys.
  Rev. A}\ }\textbf {\bibinfo {volume} {99}},\ \bibinfo {pages} {010103}
  (\bibinfo {year} {2019})}\BibitemShut {NoStop}%
\bibitem [{\citenamefont {Elitzur}\ and\ \citenamefont
  {Vaidman}(1993)}]{VE:93:FOP}%
  \BibitemOpen
  \bibfield  {author} {\bibinfo {author} {\bibfnamefont {A.}~\bibnamefont
  {Elitzur}}\ and\ \bibinfo {author} {\bibfnamefont {L.}~\bibnamefont
  {Vaidman}},\ }\href@noop {} {\bibfield  {journal} {\bibinfo  {journal}
  {Found. Phys.}\ }\textbf {\bibinfo {volume} {23}},\ \bibinfo {pages} {987}
  (\bibinfo {year} {1993})}\BibitemShut {NoStop}%
\bibitem [{\citenamefont {Kwiat}\ \emph {et~al.}(1995)\citenamefont {Kwiat},
  \citenamefont {Weinfurter}, \citenamefont {Herzog}, \citenamefont
  {Zeilinger},\ and\ \citenamefont {Kasevich}}]{KWHZK:95:PRL}%
  \BibitemOpen
  \bibfield  {author} {\bibinfo {author} {\bibfnamefont {P.}~\bibnamefont
  {Kwiat}}, \bibinfo {author} {\bibfnamefont {H.}~\bibnamefont {Weinfurter}},
  \bibinfo {author} {\bibfnamefont {T.}~\bibnamefont {Herzog}}, \bibinfo
  {author} {\bibfnamefont {A.}~\bibnamefont {Zeilinger}},\ and\ \bibinfo
  {author} {\bibfnamefont {M.~A.}\ \bibnamefont {Kasevich}},\ }\href@noop {}
  {\bibfield  {journal} {\bibinfo  {journal} {Phys. Rev. Lett.}\ }\textbf
  {\bibinfo {volume} {74}},\ \bibinfo {pages} {4763} (\bibinfo {year}
  {1995})}\BibitemShut {NoStop}%
\bibitem [{\citenamefont {Zaman}\ \emph
  {et~al.}(2019{\natexlab{a}})\citenamefont {Zaman}, \citenamefont {Shin},\
  and\ \citenamefont {Win}}]{ZSW:19:arXiv}%
  \BibitemOpen
  \bibfield  {author} {\bibinfo {author} {\bibfnamefont {F.}~\bibnamefont
  {Zaman}}, \bibinfo {author} {\bibfnamefont {H.}~\bibnamefont {Shin}},\ and\
  \bibinfo {author} {\bibfnamefont {M.~Z.}\ \bibnamefont {Win}},\ }\href@noop
  {} {\bibfield  {journal} {\bibinfo  {journal} {arXiv:1910.03200}\ } (\bibinfo
  {year} {2019}{\natexlab{a}})}\BibitemShut {NoStop}%
\bibitem [{\citenamefont {Hosten}\ \emph {et~al.}(2006)\citenamefont {Hosten},
  \citenamefont {Rakher}, \citenamefont {Barreiro}, \citenamefont {Peters},\
  and\ \citenamefont {Kwiat}}]{HRBPK:06:Nature}%
  \BibitemOpen
  \bibfield  {author} {\bibinfo {author} {\bibfnamefont {O.}~\bibnamefont
  {Hosten}}, \bibinfo {author} {\bibfnamefont {M.~T.}\ \bibnamefont {Rakher}},
  \bibinfo {author} {\bibfnamefont {J.~T.}\ \bibnamefont {Barreiro}}, \bibinfo
  {author} {\bibfnamefont {N.~A.}\ \bibnamefont {Peters}},\ and\ \bibinfo
  {author} {\bibfnamefont {P.~G.}\ \bibnamefont {Kwiat}},\ }\href@noop {}
  {\bibfield  {journal} {\bibinfo  {journal} {Nature}\ }\textbf {\bibinfo
  {volume} {439}},\ \bibinfo {pages} {949} (\bibinfo {year}
  {2006})}\BibitemShut {NoStop}%
\bibitem [{\citenamefont {Kong}\ \emph {et~al.}(2015)\citenamefont {Kong},
  \citenamefont {Ju}, \citenamefont {Huang}, \citenamefont {Wang},
  \citenamefont {Kong}, \citenamefont {Shi}, \citenamefont {Jiang},\ and\
  \citenamefont {Du}}]{KJHWKSJD:15:PRL}%
  \BibitemOpen
  \bibfield  {author} {\bibinfo {author} {\bibfnamefont {F.}~\bibnamefont
  {Kong}}, \bibinfo {author} {\bibfnamefont {C.}~\bibnamefont {Ju}}, \bibinfo
  {author} {\bibfnamefont {P.}~\bibnamefont {Huang}}, \bibinfo {author}
  {\bibfnamefont {P.}~\bibnamefont {Wang}}, \bibinfo {author} {\bibfnamefont
  {X.}~\bibnamefont {Kong}}, \bibinfo {author} {\bibfnamefont {F.}~\bibnamefont
  {Shi}}, \bibinfo {author} {\bibfnamefont {L.}~\bibnamefont {Jiang}},\ and\
  \bibinfo {author} {\bibfnamefont {J.}~\bibnamefont {Du}},\ }\href@noop {}
  {\bibfield  {journal} {\bibinfo  {journal} {Phys. Rev. Lett.}\ }\textbf
  {\bibinfo {volume} {115}},\ \bibinfo {pages} {080501} (\bibinfo {year}
  {2015})}\BibitemShut {NoStop}%
\bibitem [{\citenamefont {Noh}(2009)}]{N:09:PRL}%
  \BibitemOpen
  \bibfield  {author} {\bibinfo {author} {\bibfnamefont {T.-G.}\ \bibnamefont
  {Noh}},\ }\href@noop {} {\bibfield  {journal} {\bibinfo  {journal} {Phys.
  Rev. Lett.}\ }\textbf {\bibinfo {volume} {103}},\ \bibinfo {pages} {230501}
  (\bibinfo {year} {2009})}\BibitemShut {NoStop}%
\bibitem [{\citenamefont {Salih}(2014)}]{SH:14:PRA}%
  \BibitemOpen
  \bibfield  {author} {\bibinfo {author} {\bibfnamefont {H.}~\bibnamefont
  {Salih}},\ }\href@noop {} {\bibfield  {journal} {\bibinfo  {journal} {Phys.
  Rev. A}\ }\textbf {\bibinfo {volume} {90}},\ \bibinfo {pages} {012333}
  (\bibinfo {year} {2014})}\BibitemShut {NoStop}%
\bibitem [{\citenamefont {Itano}\ \emph {et~al.}(1990)\citenamefont {Itano},
  \citenamefont {Heinzen}, \citenamefont {Bollinger},\ and\ \citenamefont
  {Wineland}}]{IHBW:90:PRA}%
  \BibitemOpen
  \bibfield  {author} {\bibinfo {author} {\bibfnamefont {W.~M.}\ \bibnamefont
  {Itano}}, \bibinfo {author} {\bibfnamefont {D.~J.}\ \bibnamefont {Heinzen}},
  \bibinfo {author} {\bibfnamefont {J.~J.}\ \bibnamefont {Bollinger}},\ and\
  \bibinfo {author} {\bibfnamefont {D.~J.}\ \bibnamefont {Wineland}},\
  }\href@noop {} {\bibfield  {journal} {\bibinfo  {journal} {Phys. Rev. A}\
  }\textbf {\bibinfo {volume} {41}},\ \bibinfo {pages} {2295} (\bibinfo {year}
  {1990})}\BibitemShut {NoStop}%
\bibitem [{\citenamefont {Zaman}\ \emph {et~al.}(2018)\citenamefont {Zaman},
  \citenamefont {Jeong},\ and\ \citenamefont {Shin}}]{ZJS:18:SR}%
  \BibitemOpen
  \bibfield  {author} {\bibinfo {author} {\bibfnamefont {F.}~\bibnamefont
  {Zaman}}, \bibinfo {author} {\bibfnamefont {Y.}~\bibnamefont {Jeong}},\ and\
  \bibinfo {author} {\bibfnamefont {H.}~\bibnamefont {Shin}},\ }\href@noop {}
  {\bibfield  {journal} {\bibinfo  {journal} {Sci. Rep.}\ }\textbf {\bibinfo
  {volume} {8}},\ \bibinfo {pages} {14641} (\bibinfo {year}
  {2018})}\BibitemShut {NoStop}%
\bibitem [{\citenamefont {Zaman}\ \emph
  {et~al.}(2019{\natexlab{b}})\citenamefont {Zaman}, \citenamefont {Jeong},\
  and\ \citenamefont {Shin}}]{ZJS:19:SR}%
  \BibitemOpen
  \bibfield  {author} {\bibinfo {author} {\bibfnamefont {F.}~\bibnamefont
  {Zaman}}, \bibinfo {author} {\bibfnamefont {Y.}~\bibnamefont {Jeong}},\ and\
  \bibinfo {author} {\bibfnamefont {H.}~\bibnamefont {Shin}},\ }\href@noop {}
  {\bibfield  {journal} {\bibinfo  {journal} {Sci. Rep.}\ }\textbf {\bibinfo
  {volume} {9}},\ \bibinfo {pages} {11193} (\bibinfo {year}
  {2019}{\natexlab{b}})}\BibitemShut {NoStop}%
\bibitem [{Note1()}]{Note1}%
  \BibitemOpen
  \bibinfo {note} {See Supplementary Material for counterfactual implementation
  of global operations.}\BibitemShut {Stop}%
\bibitem [{\citenamefont {Huang}\ \emph {et~al.}(2004)\citenamefont {Huang},
  \citenamefont {Ren}, \citenamefont {Zhang}, \citenamefont {Duan},\ and\
  \citenamefont {Guo}}]{HXZDG:04:PRL}%
  \BibitemOpen
  \bibfield  {author} {\bibinfo {author} {\bibfnamefont {Y.-F.}\ \bibnamefont
  {Huang}}, \bibinfo {author} {\bibfnamefont {X.-F.}\ \bibnamefont {Ren}},
  \bibinfo {author} {\bibfnamefont {Y.-S.}\ \bibnamefont {Zhang}}, \bibinfo
  {author} {\bibfnamefont {L.-M.}\ \bibnamefont {Duan}},\ and\ \bibinfo
  {author} {\bibfnamefont {G.-C.}\ \bibnamefont {Guo}},\ }\href@noop {}
  {\bibfield  {journal} {\bibinfo  {journal} {Phys. Rev. Lett.}\ }\textbf
  {\bibinfo {volume} {93}},\ \bibinfo {pages} {240501} (\bibinfo {year}
  {2004})}\BibitemShut {NoStop}%
\bibitem [{\citenamefont {Huelga}\ \emph {et~al.}(2001)\citenamefont {Huelga},
  \citenamefont {Vaccaro}, \citenamefont {Chefles},\ and\ \citenamefont
  {Plenio}}]{HVCP:01:PRA}%
  \BibitemOpen
  \bibfield  {author} {\bibinfo {author} {\bibfnamefont {S.~F.}\ \bibnamefont
  {Huelga}}, \bibinfo {author} {\bibfnamefont {J.~A.}\ \bibnamefont {Vaccaro}},
  \bibinfo {author} {\bibfnamefont {A.}~\bibnamefont {Chefles}},\ and\ \bibinfo
  {author} {\bibfnamefont {M.~B.}\ \bibnamefont {Plenio}},\ }\href@noop {}
  {\bibfield  {journal} {\bibinfo  {journal} {Phys. Rev. A}\ }\textbf {\bibinfo
  {volume} {63}},\ \bibinfo {pages} {042303} (\bibinfo {year}
  {2001})}\BibitemShut {NoStop}%
\bibitem [{Note2()}]{Note2}%
  \BibitemOpen
  \bibinfo {note} {See (S44) in Supplementary Material.}\BibitemShut {Stop}%
\bibitem [{Note3()}]{Note3}%
  \BibitemOpen
  \bibinfo {note} {See (S50) in Supplementary Material.}\BibitemShut {Stop}%
\end{thebibliography}

\begin{thebibliography}{7}%
\makeatletter
\providecommand \@ifxundefined [1]{%
 \@ifx{#1\undefined}
}%
\providecommand \@ifnum [1]{%
 \ifnum #1\expandafter \@firstoftwo
 \else \expandafter \@secondoftwo
 \fi
}%
\providecommand \@ifx [1]{%
 \ifx #1\expandafter \@firstoftwo
 \else \expandafter \@secondoftwo
 \fi
}%
\providecommand \natexlab [1]{#1}%
\providecommand \enquote  [1]{``#1''}%
\providecommand \bibnamefont  [1]{#1}%
\providecommand \bibfnamefont [1]{#1}%
\providecommand \citenamefont [1]{#1}%
\providecommand \href@noop [0]{\@secondoftwo}%
\providecommand \href [0]{\begingroup \@sanitize@url \@href}%
\providecommand \@href[1]{\@@startlink{#1}\@@href}%
\providecommand \@@href[1]{\endgroup#1\@@endlink}%
\providecommand \@sanitize@url [0]{\catcode `\\12\catcode `\$12\catcode
  `\&12\catcode `\#12\catcode `\^12\catcode `\_12\catcode `\%12\relax}%
\providecommand \@@startlink[1]{}%
\providecommand \@@endlink[0]{}%
\providecommand \url  [0]{\begingroup\@sanitize@url \@url }%
\providecommand \@url [1]{\endgroup\@href {#1}{\urlprefix }}%
\providecommand \urlprefix  [0]{URL }%
\providecommand \Eprint [0]{\href }%
\providecommand \doibase [0]{https://doi.org/}%
\providecommand \selectlanguage [0]{\@gobble}%
\providecommand \bibinfo  [0]{\@secondoftwo}%
\providecommand \bibfield  [0]{\@secondoftwo}%
\providecommand \translation [1]{[#1]}%
\providecommand \BibitemOpen [0]{}%
\providecommand \bibitemStop [0]{}%
\providecommand \bibitemNoStop [0]{.\EOS\space}%
\providecommand \EOS [0]{\spacefactor3000\relax}%
\providecommand \BibitemShut  [1]{\csname bibitem#1\endcsname}%
\let\auto@bib@innerbib\@empty
\bibitem [{\citenamefont {Salih}\ \emph {et~al.}(2013)\citenamefont {Salih},
  \citenamefont {Li}, \citenamefont {Al-Amri},\ and\ \citenamefont
  {Zubairy}}]{SSLAZ:13:PRL}%
  \BibitemOpen
  \bibfield  {author} {\bibinfo {author} {\bibfnamefont {H.}~\bibnamefont
  {Salih}}, \bibinfo {author} {\bibfnamefont {Z.-H.}\ \bibnamefont {Li}},
  \bibinfo {author} {\bibfnamefont {M.}~\bibnamefont {Al-Amri}},\ and\ \bibinfo
  {author} {\bibfnamefont {M.~S.}\ \bibnamefont {Zubairy}},\ }\href@noop {}
  {\bibfield  {journal} {\bibinfo  {journal} {Phys. Rev. Lett.}\ }\textbf
  {\bibinfo {volume} {110}},\ \bibinfo {pages} {170502} (\bibinfo {year}
  {2013})}\BibitemShut {NoStop}%
\bibitem [{\citenamefont {Aharonov}\ and\ \citenamefont
  {Vaidman}(2019)}]{SAV:19:PRA}%
  \BibitemOpen
  \bibfield  {author} {\bibinfo {author} {\bibfnamefont {Y.}~\bibnamefont
  {Aharonov}}\ and\ \bibinfo {author} {\bibfnamefont {L.}~\bibnamefont
  {Vaidman}},\ }\href@noop {} {\bibfield  {journal} {\bibinfo  {journal} {Phys.
  Rev. A}\ }\textbf {\bibinfo {volume} {99}},\ \bibinfo {pages} {010103}
  (\bibinfo {year} {2019})}\BibitemShut {NoStop}%
\bibitem [{\citenamefont {Itano}\ \emph {et~al.}(1990)\citenamefont {Itano},
  \citenamefont {Heinzen}, \citenamefont {Bollinger},\ and\ \citenamefont
  {Wineland}}]{SIHBW:90:PRA}%
  \BibitemOpen
  \bibfield  {author} {\bibinfo {author} {\bibfnamefont {W.~M.}\ \bibnamefont
  {Itano}}, \bibinfo {author} {\bibfnamefont {D.~J.}\ \bibnamefont {Heinzen}},
  \bibinfo {author} {\bibfnamefont {J.~J.}\ \bibnamefont {Bollinger}},\ and\
  \bibinfo {author} {\bibfnamefont {D.}~\bibnamefont {Wineland}},\ }\href@noop
  {} {\bibfield  {journal} {\bibinfo  {journal} {Phys. Rev. A}\ }\textbf
  {\bibinfo {volume} {41}},\ \bibinfo {pages} {2295} (\bibinfo {year}
  {1990})}\BibitemShut {NoStop}%
\bibitem [{\citenamefont {Kwiat}\ \emph {et~al.}(1995)\citenamefont {Kwiat},
  \citenamefont {Weinfurter}, \citenamefont {Herzog}, \citenamefont
  {Zeilinger},\ and\ \citenamefont {Kasevich}}]{SKWHZK:95:PRL}%
  \BibitemOpen
  \bibfield  {author} {\bibinfo {author} {\bibfnamefont {P.}~\bibnamefont
  {Kwiat}}, \bibinfo {author} {\bibfnamefont {H.}~\bibnamefont {Weinfurter}},
  \bibinfo {author} {\bibfnamefont {T.}~\bibnamefont {Herzog}}, \bibinfo
  {author} {\bibfnamefont {A.}~\bibnamefont {Zeilinger}},\ and\ \bibinfo
  {author} {\bibfnamefont {M.~A.}\ \bibnamefont {Kasevich}},\ }\href@noop {}
  {\bibfield  {journal} {\bibinfo  {journal} {Phys. Rev. Lett.}\ }\textbf
  {\bibinfo {volume} {74}},\ \bibinfo {pages} {4763} (\bibinfo {year}
  {1995})}\BibitemShut {NoStop}%
\bibitem [{\citenamefont {Zaman}\ \emph {et~al.}(2018)\citenamefont {Zaman},
  \citenamefont {Jeong},\ and\ \citenamefont {Shin}}]{SZJS:18:SR}%
  \BibitemOpen
  \bibfield  {author} {\bibinfo {author} {\bibfnamefont {F.}~\bibnamefont
  {Zaman}}, \bibinfo {author} {\bibfnamefont {Y.}~\bibnamefont {Jeong}},\ and\
  \bibinfo {author} {\bibfnamefont {H.}~\bibnamefont {Shin}},\ }\href@noop {}
  {\bibfield  {journal} {\bibinfo  {journal} {Sci. Rep.}\ }\textbf {\bibinfo
  {volume} {8}},\ \bibinfo {pages} {14641} (\bibinfo {year}
  {2018})}\BibitemShut {NoStop}%
\bibitem [{\citenamefont {Zaman}\ \emph {et~al.}(2019)\citenamefont {Zaman},
  \citenamefont {Jeong},\ and\ \citenamefont {Shin}}]{SZJS:19:SR}%
  \BibitemOpen
  \bibfield  {author} {\bibinfo {author} {\bibfnamefont {F.}~\bibnamefont
  {Zaman}}, \bibinfo {author} {\bibfnamefont {Y.}~\bibnamefont {Jeong}},\ and\
  \bibinfo {author} {\bibfnamefont {H.}~\bibnamefont {Shin}},\ }\href@noop {}
  {\bibfield  {journal} {\bibinfo  {journal} {Sci. Rep.}\ }\textbf {\bibinfo
  {volume} {9}},\ \bibinfo {pages} {11193} (\bibinfo {year} {2019})}\BibitemShut
  {NoStop}%
\bibitem [{Note4()}]{Note4}%
  \BibitemOpen
  \bibinfo {note} {In the presence of the absorptive object, the H(V)-\ac
  {CQZ}$_{M,N}$ gate transforms the V (H) polarized photon to -H
  (-V).}\BibitemShut {Stop}%
\end{thebibliography}
\end{document}